\documentclass[twocolumn,aps,prc,letterpaper,superscriptaddress,nofootinbib,floatfix]{revtex4}

\usepackage{graphicx}
\usepackage{xspace}

\newcommand{\pp} {\mbox{$p$$+$$p$}\xspace}
\newcommand{\dAu} {\mbox{$d$+Au}\xspace}

\begin{document}

\title{Heavy-flavor electron-muon correlations in $p$$+$$p$ and $d$+Au collisions \\
at $\sqrt{s_{_{NN}}}$ = 200~GeV }

\newcommand{\abilene}{Abilene Christian University, Abilene, Texas 79699, USA}
\newcommand{\acadsin}{Institute of Physics, Academia Sinica, Taipei 11529, Taiwan}
\newcommand{\augie}{Department of Physics, Augustana College, Sioux Falls, South Dakota 57197, USA}
\newcommand{\banaras}{Department of Physics, Banaras Hindu University, Varanasi 221005, India}
\newcommand{\barc}{Bhabha Atomic Research Centre, Bombay 400 085, India}
\newcommand{\baruch}{Baruch College, City University of New York, New York, New York, 10010 USA}
\newcommand{\bnlcoll}{Collider-Accelerator Department, Brookhaven National Laboratory, Upton, New York 11973-5000, USA}
\newcommand{\bnlphys}{Physics Department, Brookhaven National Laboratory, Upton, New York 11973-5000, USA}
\newcommand{\caucr}{University of California - Riverside, Riverside, California 92521, USA}
\newcommand{\charlesczech}{Charles University, Ovocn\'{y} trh 5, Praha 1, 116 36, Prague, Czech Republic}
\newcommand{\chonbuk}{Chonbuk National University, Jeonju, 561-756, Korea}
\newcommand{\ciae}{Science and Technology on Nuclear Data Laboratory, China Institute of Atomic Energy, Beijing 102413, P.~R.~China}
\newcommand{\cns}{Center for Nuclear Study, Graduate School of Science, University of Tokyo, 7-3-1 Hongo, Bunkyo, Tokyo 113-0033, Japan}
\newcommand{\colorado}{University of Colorado, Boulder, Colorado 80309, USA}
\newcommand{\columbia}{Columbia University, New York, New York 10027 and Nevis Laboratories, Irvington, New York 10533, USA}
\newcommand{\czechtech}{Czech Technical University, Zikova 4, 166 36 Prague 6, Czech Republic}
\newcommand{\dapnia}{Dapnia, CEA Saclay, F-91191, Gif-sur-Yvette, France}
\newcommand{\debrecen}{Debrecen University, H-4010 Debrecen, Egyetem t{\'e}r 1, Hungary}
\newcommand{\elte}{ELTE, E{\"o}tv{\"o}s Lor{\'a}nd University, H - 1117 Budapest, P{\'a}zm{\'a}ny P. s. 1/A, Hungary}
\newcommand{\ewha}{Ewha Womans University, Seoul 120-750, Korea}
\newcommand{\fit}{Florida Institute of Technology, Melbourne, Florida 32901, USA}
\newcommand{\fsu}{Florida State University, Tallahassee, Florida 32306, USA}
\newcommand{\gsu}{Georgia State University, Atlanta, Georgia 30303, USA}
\newcommand{\hiroshima}{Hiroshima University, Kagamiyama, Higashi-Hiroshima 739-8526, Japan}
\newcommand{\ihepprot}{IHEP Protvino, State Research Center of Russian Federation, Institute for High Energy Physics, Protvino, 142281, Russia}
\newcommand{\illuiuc}{University of Illinois at Urbana-Champaign, Urbana, Illinois 61801, USA}
\newcommand{\inrras}{Institute for Nuclear Research of the Russian Academy of Sciences, prospekt 60-letiya Oktyabrya 7a, Moscow 117312, Russia}
\newcommand{\instpasczech}{Institute of Physics, Academy of Sciences of the Czech Republic, Na Slovance 2, 182 21 Prague 8, Czech Republic}
\newcommand{\isu}{Iowa State University, Ames, Iowa 50011, USA}
\newcommand{\jaea}{Advanced Science Research Center, Japan Atomic Energy Agency, 2-4 Shirakata Shirane, Tokai-mura, Naka-gun, Ibaraki-ken 319-1195, Japan}
\newcommand{\jinrdubna}{Joint Institute for Nuclear Research, 141980 Dubna, Moscow Region, Russia}
\newcommand{\jyvaskyla}{Helsinki Institute of Physics and University of Jyv{\"a}skyl{\"a}, P.O.Box 35, FI-40014 Jyv{\"a}skyl{\"a}, Finland}
\newcommand{\kek}{KEK, High Energy Accelerator Research Organization, Tsukuba, Ibaraki 305-0801, Japan}
\newcommand{\korea}{Korea University, Seoul, 136-701, Korea}
\newcommand{\kurchatov}{Russian Research Center ``Kurchatov Institute", Moscow, 123098 Russia}
\newcommand{\kyoto}{Kyoto University, Kyoto 606-8502, Japan}
\newcommand{\labllr}{Laboratoire Leprince-Ringuet, Ecole Polytechnique, CNRS-IN2P3, Route de Saclay, F-91128, Palaiseau, France}
\newcommand{\lahorelums}{Physics Department, Lahore University of Management Sciences, Lahore, Pakistan}
\newcommand{\lawllnl}{Lawrence Livermore National Laboratory, Livermore, California 94550, USA}
\newcommand{\losalamos}{Los Alamos National Laboratory, Los Alamos, New Mexico 87545, USA}
\newcommand{\lpc}{LPC, Universit{\'e} Blaise Pascal, CNRS-IN2P3, Clermont-Fd, 63177 Aubiere Cedex, France}
\newcommand{\lund}{Department of Physics, Lund University, Box 118, SE-221 00 Lund, Sweden}
\newcommand{\maryland}{University of Maryland, College Park, Maryland 20742, USA}
\newcommand{\mass}{Department of Physics, University of Massachusetts, Amherst, Massachusetts 01003-9337, USA }
\newcommand{\michigan}{Department of Physics, University of Michigan, Ann Arbor, Michigan 48109-1040, USA}
\newcommand{\muenster}{Institut fur Kernphysik, University of Muenster, D-48149 Muenster, Germany}
\newcommand{\muhlenberg}{Muhlenberg College, Allentown, Pennsylvania 18104-5586, USA}
\newcommand{\myongji}{Myongji University, Yongin, Kyonggido 449-728, Korea}
\newcommand{\nagasaki}{Nagasaki Institute of Applied Science, Nagasaki-shi, Nagasaki 851-0193, Japan}
\newcommand{\newmex}{University of New Mexico, Albuquerque, New Mexico 87131, USA }
\newcommand{\nmsu}{New Mexico State University, Las Cruces, New Mexico 88003, USA}
\newcommand{\ohio}{Department of Physics and Astronomy, Ohio University, Athens, Ohio 45701, USA}
\newcommand{\ornl}{Oak Ridge National Laboratory, Oak Ridge, Tennessee 37831, USA}
\newcommand{\orsay}{IPN-Orsay, Universite Paris Sud, CNRS-IN2P3, BP1, F-91406, Orsay, France}
\newcommand{\peking}{Peking University, Beijing 100871, P.~R.~China}
\newcommand{\pnpi}{PNPI, Petersburg Nuclear Physics Institute, Gatchina, Leningrad region, 188300, Russia}
\newcommand{\riken}{RIKEN Nishina Center for Accelerator-Based Science, Wako, Saitama 351-0198, Japan}
\newcommand{\rikjrbrc}{RIKEN BNL Research Center, Brookhaven National Laboratory, Upton, New York 11973-5000, USA}
\newcommand{\rikkyo}{Physics Department, Rikkyo University, 3-34-1 Nishi-Ikebukuro, Toshima, Tokyo 171-8501, Japan}
\newcommand{\saispbstu}{Saint Petersburg State Polytechnic University, St. Petersburg, 195251 Russia}
\newcommand{\saopaulo}{Universidade de S{\~a}o Paulo, Instituto de F\'{\i}sica, Caixa Postal 66318, S{\~a}o Paulo CEP05315-970, Brazil}
\newcommand{\seoulnat}{Seoul National University, Seoul, Korea}
\newcommand{\stonybrkc}{Chemistry Department, Stony Brook University, SUNY, Stony Brook, New York 11794-3400, USA}
\newcommand{\stonycrkp}{Department of Physics and Astronomy, Stony Brook University, SUNY, Stony Brook, New York 11794-3400, USA}
\newcommand{\subatech}{SUBATECH (Ecole des Mines de Nantes, CNRS-IN2P3, Universit{\'e} de Nantes) BP 20722 - 44307, Nantes, France}
\newcommand{\tenn}{University of Tennessee, Knoxville, Tennessee 37996, USA}
\newcommand{\titech}{Department of Physics, Tokyo Institute of Technology, Oh-okayama, Meguro, Tokyo 152-8551, Japan}
\newcommand{\tsukuba}{Institute of Physics, University of Tsukuba, Tsukuba, Ibaraki 305, Japan}
\newcommand{\vandy}{Vanderbilt University, Nashville, Tennessee 37235, USA}
\newcommand{\waseda}{Waseda University, Advanced Research Institute for Science and Engineering, 17 Kikui-cho, Shinjuku-ku, Tokyo 162-0044, Japan}
\newcommand{\weizmann}{Weizmann Institute, Rehovot 76100, Israel}
\newcommand{\wigner}{Institute for Particle and Nuclear Physics, Wigner Research Centre for Physics, Hungarian Academy of Sciences (Wigner RCP, RMKI) H-1525 Budapest 114, POBox 49, Budapest, Hungary}
\newcommand{\yonsei}{Yonsei University, IPAP, Seoul 120-749, Korea}
\affiliation{\abilene}
\affiliation{\acadsin}
\affiliation{\augie}
\affiliation{\banaras}
\affiliation{\barc}
\affiliation{\baruch}
\affiliation{\bnlcoll}
\affiliation{\bnlphys}
\affiliation{\caucr}
\affiliation{\charlesczech}
\affiliation{\chonbuk}
\affiliation{\ciae}
\affiliation{\cns}
\affiliation{\colorado}
\affiliation{\columbia}
\affiliation{\czechtech}
\affiliation{\dapnia}
\affiliation{\debrecen}
\affiliation{\elte}
\affiliation{\ewha}
\affiliation{\fit}
\affiliation{\fsu}
\affiliation{\gsu}
\affiliation{\hiroshima}
\affiliation{\ihepprot}
\affiliation{\illuiuc}
\affiliation{\inrras}
\affiliation{\instpasczech}
\affiliation{\isu}
\affiliation{\jaea}
\affiliation{\jinrdubna}
\affiliation{\jyvaskyla}
\affiliation{\kek}
\affiliation{\korea}
\affiliation{\kurchatov}
\affiliation{\kyoto}
\affiliation{\labllr}
\affiliation{\lahorelums}
\affiliation{\lawllnl}
\affiliation{\losalamos}
\affiliation{\lpc}
\affiliation{\lund}
\affiliation{\maryland}
\affiliation{\mass}
\affiliation{\michigan}
\affiliation{\muenster}
\affiliation{\muhlenberg}
\affiliation{\myongji}
\affiliation{\nagasaki}
\affiliation{\newmex}
\affiliation{\nmsu}
\affiliation{\ohio}
\affiliation{\ornl}
\affiliation{\orsay}
\affiliation{\peking}
\affiliation{\pnpi}
\affiliation{\riken}
\affiliation{\rikjrbrc}
\affiliation{\rikkyo}
\affiliation{\saispbstu}
\affiliation{\saopaulo}
\affiliation{\seoulnat}
\affiliation{\stonybrkc}
\affiliation{\stonycrkp}
\affiliation{\subatech}
\affiliation{\tenn}
\affiliation{\titech}
\affiliation{\tsukuba}
\affiliation{\vandy}
\affiliation{\waseda}
\affiliation{\weizmann}
\affiliation{\wigner}
\affiliation{\yonsei}
\author{A.~Adare} \affiliation{\colorado}
\author{S.~Afanasiev} \affiliation{\jinrdubna}
\author{C.~Aidala} \affiliation{\mass} \affiliation{\michigan}
\author{N.N.~Ajitanand} \affiliation{\stonybrkc}
\author{Y.~Akiba} \affiliation{\riken} \affiliation{\rikjrbrc}
\author{H.~Al-Bataineh} \affiliation{\nmsu}
\author{J.~Alexander} \affiliation{\stonybrkc}
\author{A.~Angerami} \affiliation{\columbia}
\author{K.~Aoki} \affiliation{\kyoto} \affiliation{\riken}
\author{N.~Apadula} \affiliation{\stonycrkp}
% ...
\collaboration{PHENIX Collaboration} \noaffiliation

\date{\today}


\begin{abstract}

We report $e^\pm-\mu^\mp$ pair yield from charm decay measured between 
midrapidity electrons ($|\eta|<0.35$ and $p_T>0.5$~GeV/$c$) and forward 
rapidity muons ($1.4<\eta<2.1$ and $p_T>1.0$~GeV/$c$) as a function of 
$\Delta\phi$ in both $p$$+$$p$ and in $d$+Au collisions at 
$\sqrt{s_{_{NN}}}=200$~GeV. Comparing the $p$$+$$p$ results with several 
different models, we find the results are consistent with a total charm 
cross section $\sigma_{c\bar{c}} =$ 538 $\pm$ 46 (stat) $\pm$ 
197~(data~syst) $\pm$ 174~(model~syst) $\mu$b. These generators also 
indicate that the back-to-back peak at $\Delta\phi = \pi$ is dominantly 
from the leading order contributions (gluon fusion), while higher order 
processes (flavor excitation and gluon splitting) contribute to the yield 
at all $\Delta\phi$. We observe a suppression in the pair yield per 
collision in $d$+Au. We find the pair yield suppression factor for 
$2.7<\Delta\phi<3.2$ rad is $J_{dA}$ = 0.433 $\pm$ 0.087 (stat) $\pm$ 
0.135 (syst), indicating cold nuclear matter modification of $c\bar{c}$ 
pairs.

\end{abstract}


\maketitle

\section{Introduction} \label{sec:intro}

The study of open heavy flavor production in relativistic $p(d)+A$ 
collisions is sensitive to different kinds of strong interaction physics. 
Because the leading order (LO) production mechanism is gluon 
fusion~\cite{Brambilla:2010cs}, open heavy flavor production rates are 
directly related to modification of the gluon parton distribution function 
(PDF), i.e.~shadowing or saturation~\cite{Gelis:2010nm}. Also, the initial 
and/or final state partons can scatter and lose energy in the cold nuclear 
medium~\cite{Wang:2001ifa,Frankfurt:2007rn,Vitev:2007ve}, thereby 
modifying and producing a nuclear modification of open heavy flavor 
production.  Recently, the possibility of flow even in small collision 
systems such as $p(d)+A$ has raised the question of modified charm 
momentum distributions \cite{Sickles:2013yna}.

Modification to heavy quark production rates and kinematics in \dAu 
~collisions at the Relativistic Heavy Ion Collider (RHIC) is well 
established. Electron production from open heavy flavor decay is 
enhanced~\cite{Adare:2012qb}, while $J/\psi$ production is 
suppressed~\cite{Adare:2012qf} at midrapidity. At positive rapidity, 
defined with the positive $z$ axis as the direction of the deuteron, there 
is a suppression of heavy-flavor decay muons~\cite{Adare:2013lkk} and a 
larger suppression of $J/\psi$~\cite{Adare:2012qf}.

While $e$--$\mu$ correlations from open heavy flavor decays have not been 
published at RHIC to date, correlations involving light flavor hadrons 
have shown modification in \dAu collisions at RHIC. A suppression has been 
observed of positive rapidity $\pi^0$ mesons associated with midrapidity 
trigger hadrons, especially in the back-to-back peak at $\Delta\phi = 
\pi$, indicating $2\rightarrow2$ scatterings~\cite{Adare:2011sc}.  This 
suppression increases as $x$, the fraction of the nucleon momentum carried 
by the gluon, decreases.  These results are in quantitative agreement with 
energy loss models~\cite{Kang:2011bp} and saturation 
models~\cite{Stasto:2011ru,JalilianMarian:2012bd,Lappi:2012nh}.

This paper presents measurements of azimuthal correlations of 
electron-muon pairs produced from heavy flavor decays, primarily 
$c\bar{c}$, in \pp and \dAu collisions using the PHENIX detector at 
RHIC. The heavy-flavor $e$--$\mu$ correlations are free of backgrounds from 
other sources that contribute to other dilepton analyses ($e^+e^-$ or 
$\mu^+\mu^-$), such as resonance decay and Drell-Yan. While analysis of 
dilepton mass and $p_T$ provides a way to separate charm and bottom 
contributions, the azimuthal correlations have an important advantage for 
studying the charm production process. The leading order production, 
$gg\rightarrow Q\bar{Q}$ and $q\bar{q}\rightarrow Q\bar{Q}$, will produce 
back-to-back open heavy flavor pairs that can semileptonically decay and 
produce azimuthally correlated $e$--$\mu$ pairs. Next-to-leading order 
(NLO) processes like flavor excitation and gluon splitting produce much less 
correlated $Q\bar{Q}$ and thus much less correlated $e$--$\mu$ pairs. 
Therefore, modification to different portions of the azimuthal 
correlations can be attributed to modifications of $c\bar{c}$ pairs from 
different production mechanisms. In energy loss models such as 
Ref.~\cite{Kang:2011bp}, a broadening of the back-to-back azimuthal 
correlation should accompany a suppression of the peak due to the multiple 
scattering that the incoming gluons and/or the outgoing $c\bar{c}$ undergo 
in the cold nuclear medium.

This paper is organized as follows. The PHENIX detector is outlined in 
Section~\ref{sec:phenix}. Section~\ref{sec:method} describes the details 
of the method used to measure the correlations, the background subtraction 
method, and the tests of the method. Section~\ref{sec:results} presents 
the results in \pp and compares them to Monte Carlo models. The \dAu 
results are presented and compared to the \pp results in 
Section~\ref{sec:dAuresults}. Conclusions are given in 
Section~\ref{sec:end}.

\section{PHENIX Experiment}\label{sec:phenix}

\begin{figure}[tbh]
\includegraphics[width=1.0\linewidth]{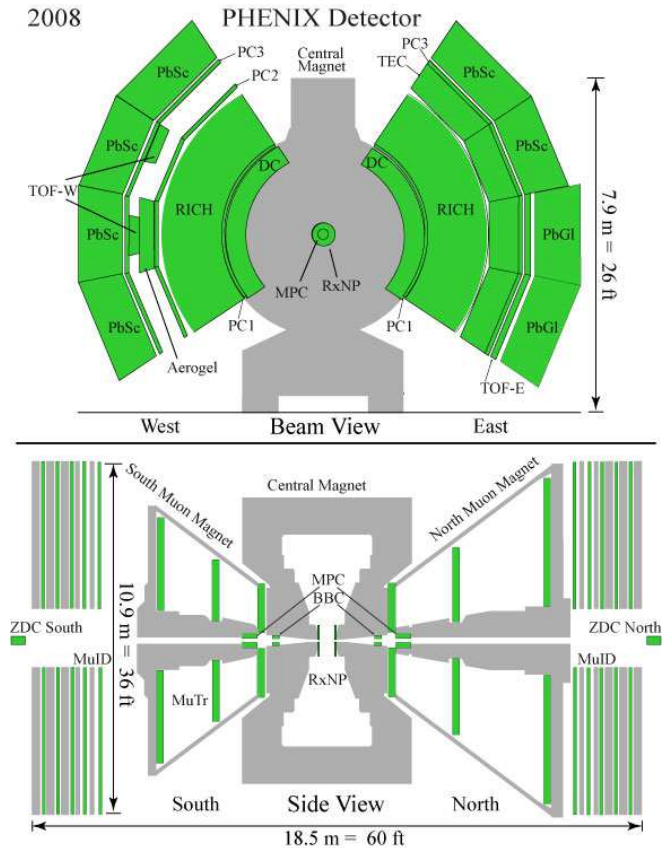}
\caption{(Color online) 
A schematic view of the PHENIX detector during the 2008 \dAu data taking. 
(a) beam view of the central spectrometer arms.  (b) 
longitudinal view including the global event and 
triggering detectors, as well as the muon spectrometer arms. The 
configurations of the central spectrometer and muon arms were the same for 
the 2006 \pp data taking.
}
\label{fig:phenix}
\end{figure}

The PHENIX detector at RHIC is multi-purposed and optimized for precision 
measurements of electromagnetic probes for relativistic hadronic and heavy 
ion collisions.  A complete overview of the detector can be found in 
Ref.~\cite{Adcox:2003zm}.  The data presented here are from 2006 \pp and 
2008 \dAu data taking at RHIC.  Figure~\ref{fig:phenix} shows a schematic 
of the detector during those years. This analysis uses the central 
spectrometer arms for electron detection and the forward rapidity muon 
spectrometer arms, labeled North and South in Fig.~\ref{fig:phenix}, for 
muon identification.  For the 2008 \dAu collisions, the deuteron beam 
moves toward the North arm, which defines positive rapidity for both \pp 
and \dAu. The forward produced muons come from a high-$x$ parton in the 
deuteron interacting with a low-$x$ parton in the gold. {\sc 
pythia}\cite{Sjostrand:2006za} indicates that the average $x$ of a parton 
producing a heavy flavor muon from 1$<p_T^{\mu}<6$~GeV/$c$ in the forward 
muon spectrometer is about $5\times10^{-3}$. This analysis focuses only on 
the muons measured in the North arm utilizing the deuteron beam as a probe 
of low-$x$ partons in the gold nucleus.

The central spectrometer comprises two arms subtending $\pi/2$ in azimuth 
and covering $|\eta|<0.35$. Charged tracks are measured using a drift 
chamber (DC) and a set of multi-wire proportional chambers with pad 
readout (PC1 and PC3). The DC measures the bend angle in the $r-\phi$ 
plane due to a central magnetic field directed along the beam axis. PC1 is 
used to measure the longitudinal coordinate of the track. These tracks are 
then projected into PC3, where a hit is required to ensure high track 
quality. The momentum resolution of the tracks in this data is 
$\delta{p/p}=1.10$\%$\oplus$1.16\%$p$, where $p$ is the total momentum 
measured in GeV/$c$. Electrons can be identified from associated hits in 
the Ring Imaging \v{C}erenkov (RICH) detector and the Electromagnetic 
Calorimeters (EMCal). Electrons above 17 MeV/$c$ passing through the 
CO$_2$-filled RICH will emit \v{C}erenkov radiation. The EMCal comprises 
eight sectors, six of lead-scintillator and two of lead-glass, used to 
collect the energy from electron and photon showers. The nominal energy 
resolution for the lead-scintillator and lead-glass is 
8.1\%$\pm\sqrt{E{\rm [GeV]}}\oplus$2.1\% and 6.0\%$\pm\sqrt{E{\rm 
[GeV]}}\oplus$0.9\%\cite{Aphecetche:2003zr}, respectively.

The North muon spectrometer is located at $1.2<\eta<2.4$ and covers 2$\pi$ 
in azimuth. The spectrometer measures tracks in the muon tracker (MuTr) 
and the muon identifier (MuID). Prior to entering the muon arm, particles 
pass through approximately 20~cm of copper and 60~cm of iron. Particles 
that are not absorbed pass through the MuTr, which comprise three 
stations of cathode strip chambers with multiple ionization regions and 
located inside a radial magnetic field. After the MuTr, particles pass 
through the MuID, which comprises five alternating steel absorbers and 
MuID detector planes, called gaps, with Iarocci tubes. MuID roads 
reconstructed from MuID hits are projected back to MuTr tracks and to the 
measured vertex to provide the complete information for a track through 
the spectrometer.

Trigger and global event characterization in \pp and \dAu are provided 
by the beam-beam counter (BBC). The BBC is a set of 64 hexagonal \v{C}erenkov 
counters located from $3.0<|\eta|<3.9$ and covering full azimuth. The 
vertex of the collision along the beam line ($z_{\rm vtx}$) is determined by 
the time difference between the BBCs on either side of the collision 
region. The minimum bias (MB) trigger requires that there is at least one 
hit in each of the BBCs. From Vernier scans and verified by Monte Carlo 
studies, the BBC MB trigger is sensitive to 55$\pm$5\% of the 
\pp inelastic cross section and 88$\pm$4\% of the \dAu inelastic 
cross section\cite{Adare:2013nff}. The trigger used for this analysis is a 
combination of the BBC trigger and a deep muon trigger. The deep muon 
trigger requires three or more MuID gaps with a signal in both the $x$ and 
$y$ direction tubes and that the last pair of hits be in the last (5th 
gap) or next to last gap (4th gap).

After quality cuts and requiring a vertex within 25~cm of the $z$=0 
vertex, an integrated luminosity of 2.1 pb$^{-1}$ in \pp and a 
\pp-equivalent of 7.7 pb$^{-1}$ in \dAu was sampled.

\section{Analysis}\label{sec:method}
The primary goal of this analysis is to identify 
\begin{equation}
p+p(d+{\rm Au}) \rightarrow c\bar{c} + X \rightarrow e^{\pm}\mu^{\mp} + X,
\end{equation}
where the opposite sign electron-muon pair is from the $c\bar{c}$ pair 
decay.

\subsection{Particle Identification}\label{sec:pid}

\subsubsection{Muon Identification}

Only muon candidates with $p_T>1$~GeV/$c$ are used in the analysis 
because real muons with total momentum less than about 2.7~GeV/$c$ are 
stopped in the muon arm before reaching the 5th (and last) gap. Single 
muon candidates are constructed from MuID roads projected and matched to 
MuTr tracks. Cuts on MuID roads and MuTr tracks are designed to reject 
hadrons that mimic a muon signal and to reject tracks that did not 
originate from the collision vertex. For the MuID roads, at least three of 
five gaps with $x-y$ hit information are required, including a pair of 
hits in the 5th gap. These MuID roads must project back near the nominal 
vertex position. Those muons that do not typically come from beam-related 
backgrounds. For the MuTr tracks, cuts that reject hadrons are detailed in 
Ref.~\cite{Adler:2006yu}. The MuID roads are then projected and matched to 
MuTr tracks at the 1st MuID gap. An identified muon candidate is the 
closest MuTr track that matches a MuID road within at least 10$^\circ$ in 
slope and 10~cm in distance. Muon candidates are further restricted to 
$1.4<\eta<2.1$. During both the \pp and \dAu data taking periods, there were 
backgrounds primarily from beam-related particles interacting with 
material in the accelerator upstream of PHENIX. Collimators were used in 
the accelerator to reduce this background but it was not totally 
eliminated. Restricting the $\eta$ range of the muon candidates helped 
minimize this background.

\subsubsection{Electron Identification}

Electrons with $p_T>0.5$~GeV/$c$ are identified by matching a track in 
DC, PC1, and PC3 to a signal in the RICH and a cluster in the EMCal. The relevant 
details on measuring electrons in PHENIX are given in 
Ref.~\cite{Adare:2010de}. For this analysis, the projected track must 
match within 3$\sigma$ in position to a cluster in the EMCal. Clusters are 
also required to have a matching profile, when compared to an 
electromagnetic shower shape profile at the measured energy. Once a track 
matches both the RICH and the EMCal, an $E/p$ cut is applied, where it is 
required that the energy measured in the the EMCal $E$ be approximately 
equal to the reconstructed track momentum $p$. This is sufficient to 
remove most combinatorial matches and background from real electrons 
resulting from long-lived particle decays occurring near the DC, which 
have mismeasured momentum. A cut of -2$\sigma$ to +3$\sigma$ from the mean 
$E/p$ in the \pp data and -1.5$\sigma$ to +3$\sigma$ from the mean in the 
\dAu data is applied. The asymmetry of the cuts is due to the dominance of 
backgrounds below 2 or 1.5$\sigma$ of the mean. The tighter cut in the 
\dAu data was necessary because of the increased background from the 
hadron blind detector (HBD) support material not present during 2006 
data taking.

\subsection{Acceptance and Efficiencies}\label{sec:effacc}

After particle identification cuts have been applied to an event, all 
pairs of identified electrons and muons are formed in each of the four 
charge-sign combinations. The fully corrected invariant-pair yield, 
calculated for each sign combination, is \cite{Adler:2005ad}
\begin{eqnarray}\label{eq:pairyield}
\frac{d^3N}{dy^{\mu}dy^ed\Delta\phi} = \frac{c}{N^{\rm MB}_{\rm evt}\Delta y^e \Delta y^\mu\Delta\phi^{bin}}\times \nonumber \\
\quad \frac{\int d\Delta\phi~{\rm Mix}(\Delta\phi)}{2\pi} \frac{N^{e\mu}(\Delta\phi)}{{\rm Mix^{e\mu}}(\Delta\phi,\epsilon^e,\epsilon^\mu)}
\end{eqnarray}

where $N^{\rm MB}_{\rm evt}$ is the number of sampled BBC triggered events, $c$ is 
the MB trigger bias accounting for events missed by the BBC 
trigger\cite{Adare:2013nff}, $\Delta y^e$ and $\Delta y^\mu$ are the 
rapidity ranges of the electrons and muons, respectively, 
$N^{e\mu}(\Delta\phi)$ is the inclusive electron-muon pair yield, and 
${\rm Mix}^{e\mu}(\Delta\phi,\epsilon^e,\epsilon^\mu)$ is the 
mixed-event electron-muon pair distribution.  The two-particle acceptance 
and efficiency is corrected by the mixed-event technique, where electrons 
from one event are paired with muons from a different event. Pools of 
inclusive electrons and muons are kept in bins 2.5~cm-wide $z$ vertex 
bins, and, in the case of \dAu, 10\%-wide centrality bins.  When mixing 
events, the pair distribution is weighted by the $y$- and $\phi$-averaged 
efficiency of each particle, $\epsilon^e$ and $\epsilon^\mu$.

Both $\epsilon^e$ and $\epsilon^\mu$ were determined by generating single 
electrons and single muons with a flat distribution in $p_T$, $\phi$, 
$|y^e|<0.5$ or $1.4<y^\mu<2.2$ and collisions $z$-vertex location and 
running them through a {\sc geant}-3 simulation of the PHENIX detector. The 
output was subjected to the same analysis cuts applied to the data. The 
efficiency is defined as the ratio of particles reconstructed through the 
analysis to the number simulated. These simulations demonstrated that 
$\epsilon^e$ and $\epsilon^\mu$ are independent of the $z$-position of the 
event vertex, $\epsilon_e$ is independent of $\eta$ and $\epsilon_\mu$ has 
a slight $\eta$-dependence. Pair yields are reported with the average 
pseudorapidity $\langle\eta^\mu\rangle$, which include the 
$\eta$-dependence of both single inclusive muons and the single particle 
efficiency.

\subsection{Background Subtraction}\label{sec:subtr}

Inclusive muon and electron candidates come from both heavy- and 
light-flavor decays and from misidentified hadrons.  The fully-corrected 
inclusive electron-muon pair yield for each sign combinations can be 
written as

\begin{eqnarray}\label{eq:basecorr}
N^{e\mu}(\Delta\phi) & = & N^{e\mu}_{H}(\Delta\phi) + N^{e\mu}_{LH}(\Delta\phi) + N^{e\mu}_{L}(\Delta\phi).
\end{eqnarray}

Here $N^{e\mu}$ indicates the fully-corrected inclusive pair yield defined 
in Eq.~\ref{eq:pairyield}; $N^{e\mu}_{H}(\Delta\phi)$ is the 
fully-corrected pair yield produced from a heavy flavor pair decay; 
$N^{e\mu}_{LH}(\Delta\phi)$ is the fully-corrected pair yield from 
correlating a heavy flavor decay product with a light flavor decay 
product; and $N^{e\mu}_{L}(\Delta\phi)$ is the fully-corrected pair yield 
from correlating pairs of light-flavor decay products or misidentified 
hadrons. Pairs from the semileptonic decay of a $c\bar{c}$ pair have 
opposite signs. Eq.~\ref{eq:basecorr} can be decomposed into its like- and 
unlike-sign pieces:

\begin{eqnarray}\label{eq:signedocrr}
N^{e\mu}_{{\rm like}}(\Delta\phi) & = & N^{e\mu}_{LH,{\rm like}}(\Delta\phi) + N^{e\mu}_{L,{\rm like}}(\Delta\phi) \nonumber \\
N^{e\mu}_{{\rm unlike}}(\Delta\phi) & = & N^{e\mu}_{H,{\rm unlike}}(\Delta\phi) + N^{e\mu}_{LH,{\rm unlike}}(\Delta\phi) + \nonumber \\
& & N^{e\mu}_{L,{\rm unlike}}(\Delta\phi).
\end{eqnarray}

While semileptonic decays of $b\bar{b}$ can also produce both like- and 
unlike-sign $e$--$\mu$ signals, in this analysis, {\sc pythia} indicates 
that only about 1\% of the final heavy-flavor $e$--$\mu$ pair yield is 
from $b\bar{b}$ and is neglected. If we assume muon (electron) candidates 
from light flavors are not charge-correlated with electron (muon) 
candidates from light flavors, then
\begin{equation}\label{eqLikeUnlike1}
N^{e\mu}_{L,{\rm like}}(\Delta\phi) = N^{e\mu}_{L,{\rm unlike}}(\Delta\phi).
\end{equation}
If only one of the pair is from heavy flavor, then again, we assume they 
are not charge correlated, and
\begin{equation}\label{eqLikeUnlike2}
 N^{e\mu}_{LH,{\rm like}}(\Delta\phi) = N^{e\mu}_{LH,{\rm unlike}}(\Delta\phi). 
\end{equation}
Therefore, the heavy flavor $e$--$\mu$ signal distributions is the 
difference between the unlike-sign and the like-sign inclusive 
correlations:
\begin{eqnarray}\label{eq:subtrcorr}
N^{e\mu}_{H}(\Delta\phi) & = & N^{e\mu}_{{\rm unlike}}(\Delta\phi) - N^{e\mu}_{{\rm like}}(\Delta\phi).
\end{eqnarray}

\begin{figure}
\includegraphics[width=1.0\linewidth]{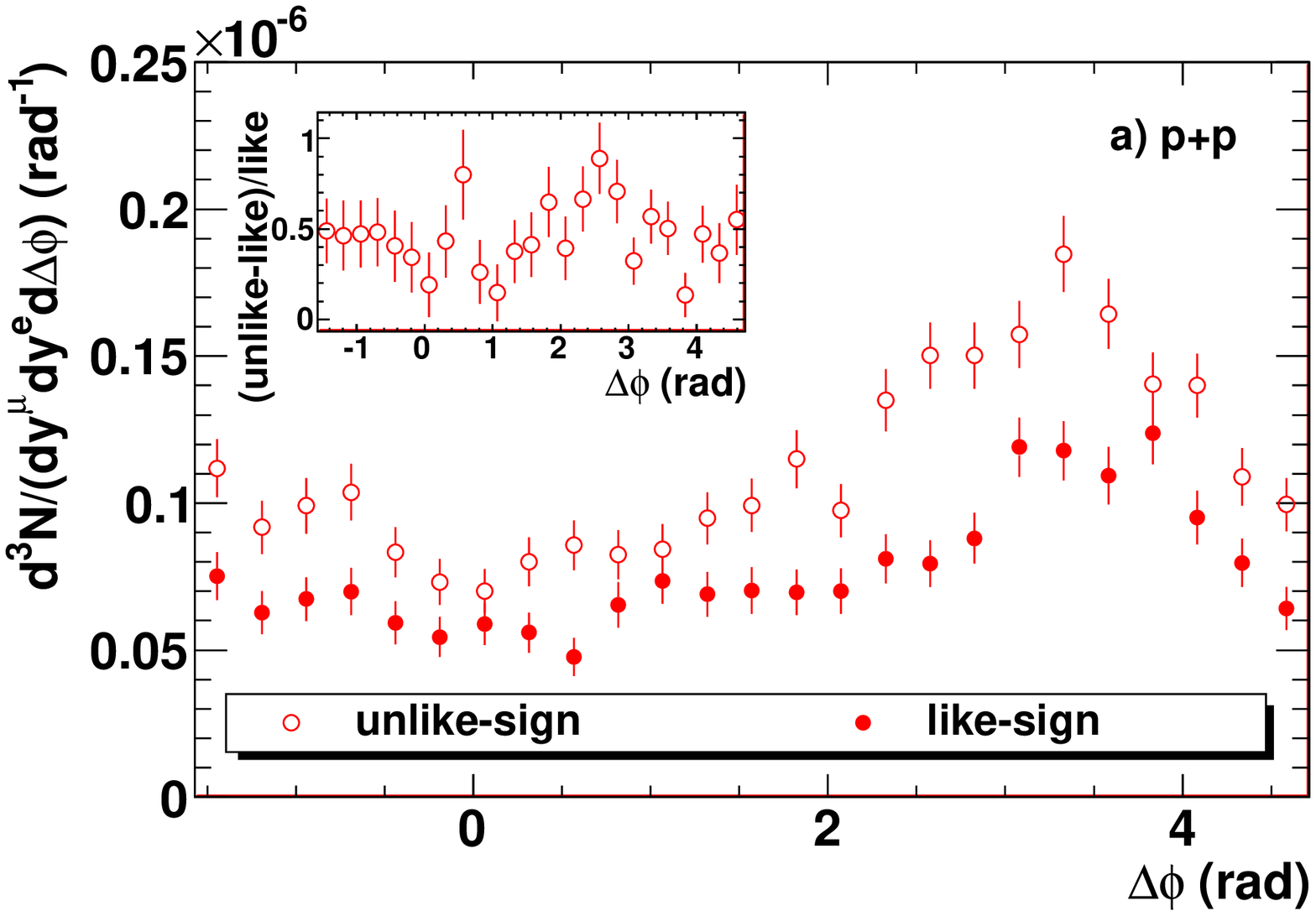}
\includegraphics[width=1.0\linewidth]{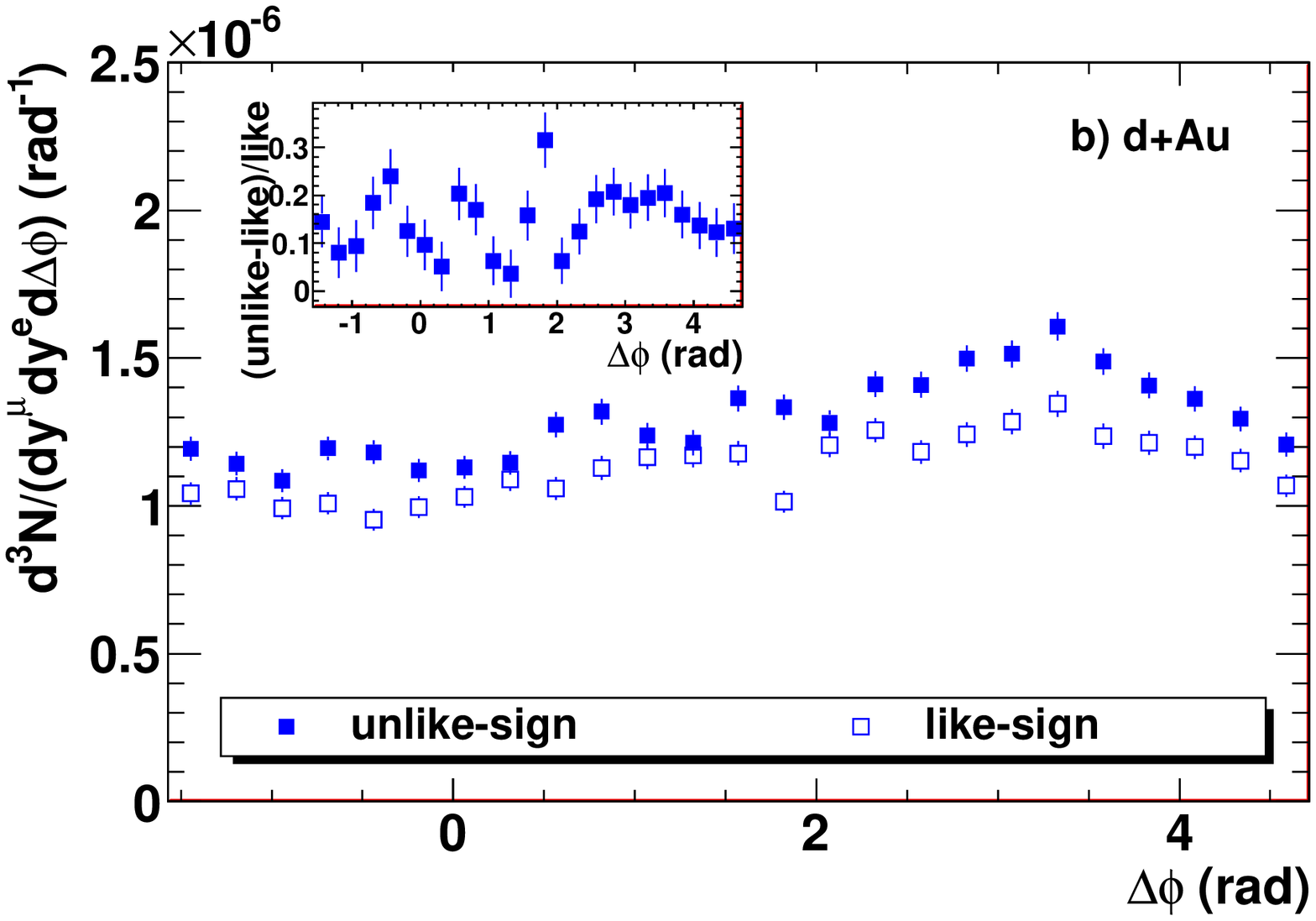}
\caption{(Color online) 
The fully-corrected inclusive like-sign ($e^{\pm}-\mu^{\pm}$) and 
unlike-sign ($e^{\pm}-\mu^{\mp}$) distributions for (a) \pp and (b) \dAu, 
as a function of $\Delta\phi$. The inset shows the unlike-like difference 
divided by the like-sign distribution, which is the heavy flavor 
signal-to-background in the inclusive unlike-sign distribution.
}
\label{fig:sb}
\end{figure}

Figure~\ref{fig:sb} shows the fully-corrected inclusive like-sign 
($N^{e\mu}_{{\rm like}}(\Delta\phi)$) and unlike-sign 
($N^{e\mu}_{{\rm unlike}}(\Delta\phi)$) $e$--$\mu$ pair distributions in 
\pp and \dAu. The inset figures show the signal-to-background 
distributions given the assumptions above. 

We have checked the like-sign subtraction method using {\sc pythia} leading 
order quantum chromodynamics (QCD) events. With all events containing a heavy quark in the final 
state removed, the pair yields as a function of $\Delta\phi$ for like-sign 
and unlike-sign electron-muon pairs were the same within 3\% over 
all $\Delta\phi$.

While this corroborates the basic idea of the subtraction, the assumption 
was further tested with data. In the following sections we detail the 
results of different methods to tag electrons and muons from light flavor 
decay to examine the validity of Eq.~\ref{eq:subtrcorr} and to quantify 
the systematic uncertainty of the method. The general method is to use a 
sample of single electrons paired with single muons, where one or both are 
likely from light-hadron decays. If the method is correct, the like-sign 
subtraction should produce no correlation at all. If there are 
statistically significant correlations after like-sign subtraction, these 
are subtracted from the final $e$--$\mu$ pair yield and uncertainties on 
the residual correlation strength are propagated as a systematic 
uncertainty on the final pair yield. If no statistically significant yield 
is found after like-sign subtraction, the statistical uncertainty on the 
zero yield is propagated as the systematic uncertainty.

\subsubsection{Correlations between inclusive electrons and punch-through 
hadrons that fake single muons}\label{sec:pppunch-through}

\begin{figure}[thb]
\includegraphics[width=1.0\linewidth]{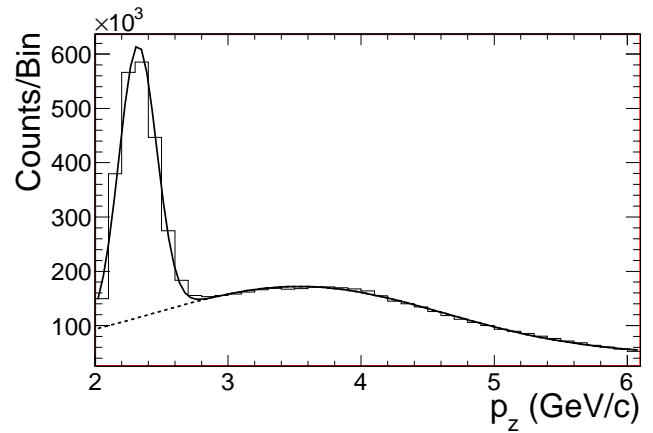}
\caption{
The distribution of $p_z$ for tracks that stop in the next-to-last MuID 
gap (4th gap). The peak at lower $p_z$ is due to muons, while the broad 
distribution is from hadrons that punch through the absorber to the 4th 
gap. The solid line is a two-Gaussian fit to this distribution with the 
solid line indicating the hadronic background in the muon peak region.
}
\label{fig:pz}
\end{figure}

One source of background to the single muons is from hadrons that 
penetrate to the 5th gap, called punch-through hadrons. After single 
particle cuts there is some small fraction (roughly 1 out of every 
250~\cite{Hornback:2008sba}) of candidate tracks with $p_T>1$~GeV/$c$ 
that are hadrons that punch through. While this represents an irreducible 
background to the single muons, we can obtain a clean sample of hadrons 
that punch through and stop in the 4th gap of the MuID. Fig.~\ref{fig:pz} 
shows the $p_z$ distribution of muon candidates that stop in the 4th gap. 
The peak at 2.3~GeV is composed of muons that have insufficient energy to 
penetrate further. The broader portion of the distribution comprises 
light hadrons that are not stopped by the upstream absorber materials but 
are subsequently absorbed in the steel just after the 4th gap, thus not 
leaving a hit in the 5th gap. We identify punch-through hadrons as having 
stopped in the 4th gap with $p_z$ larger than 3~GeV.

\begin{figure}[thb]
\includegraphics[width=1.0\linewidth]{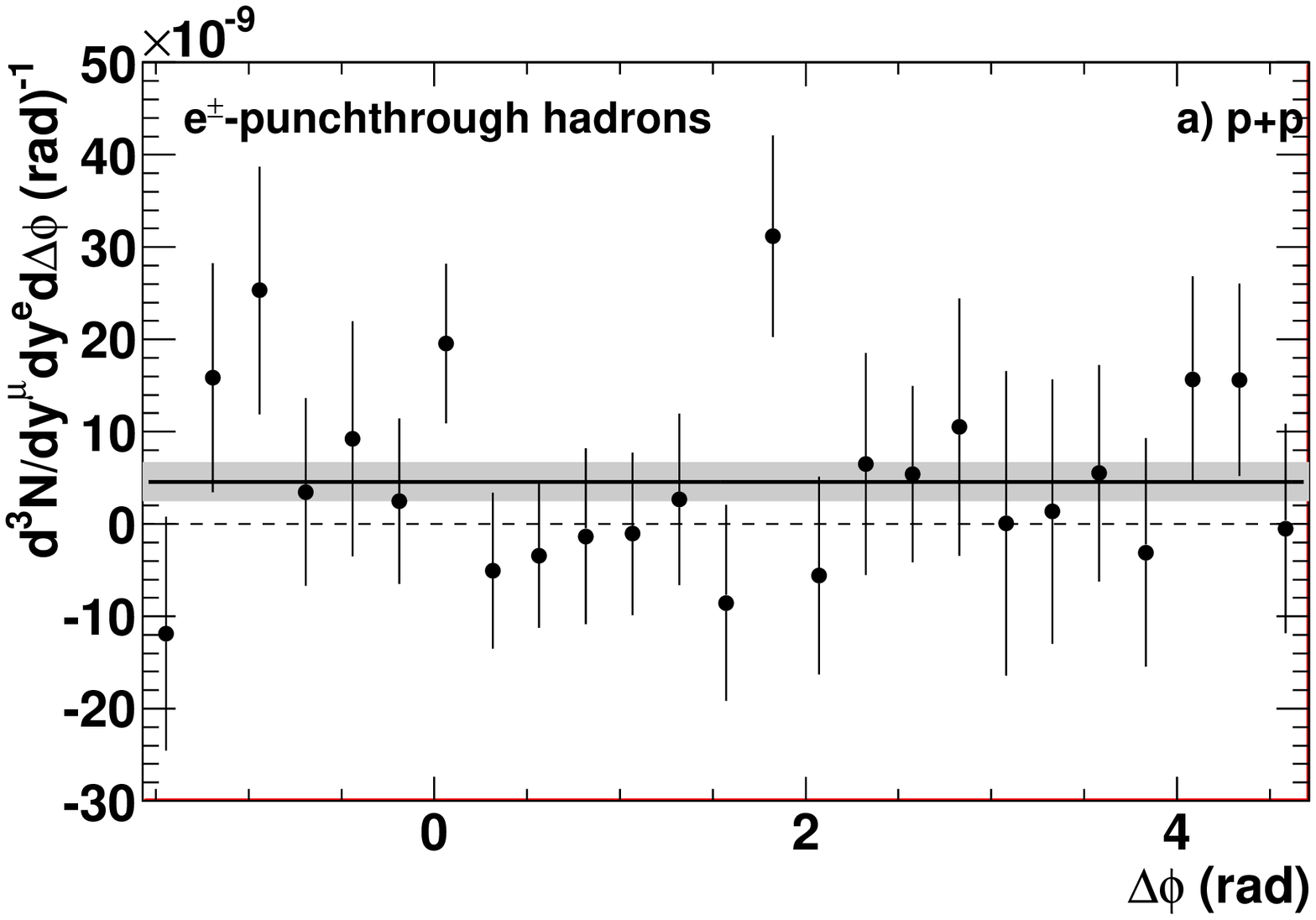}
\includegraphics[width=1.0\linewidth]{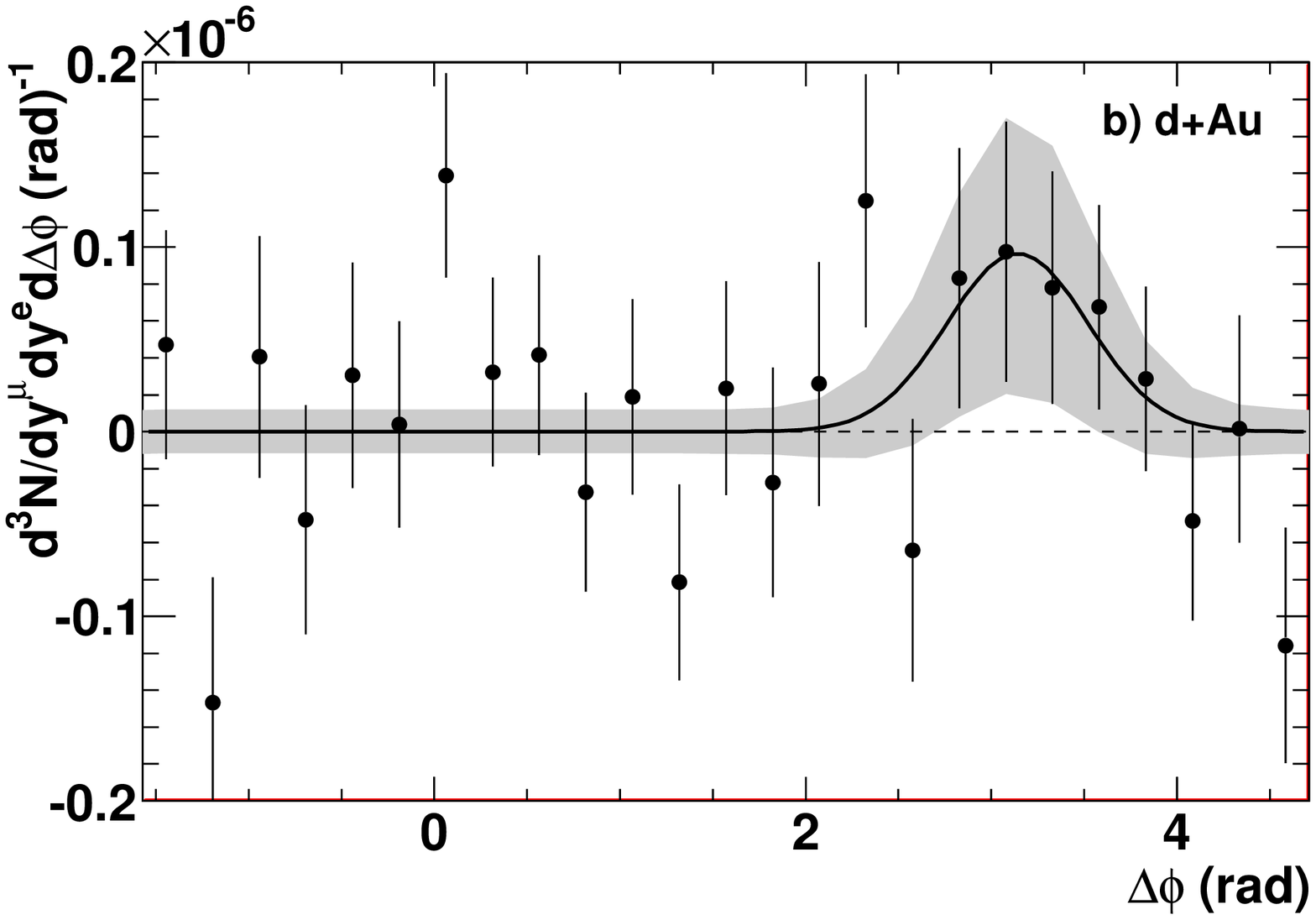}
\caption{
The fully-corrected like-sign-subtracted electron plus punch-through 
hadron pair yield in (a) \pp and (b) \dAu collisions. The line indicates 
the fitted yield that is removed from the inclusive electron-muon pair 
correlation. The shaded band indicates the fit uncertainty that is 
propagated as a systematic uncertainty in the final pair yield. In \pp the 
fit is is a flat line with $\chi^2$/NDF = 22.7/24. In \dAu it is a flat 
line and a Gaussian centered at $\pi$ with $\chi^2$/NDF = 
26.3/22.}
\label{fig:punch-throughcorr}
\end{figure}

Fig.~\ref{fig:punch-throughcorr} shows the fully corrected like-sign 
subtracted pair yield of central-arm electrons and the punch-through 
hadrons in the muon arms for both \pp and \dAu collisions. If both the 
like- and unlike-sign pair yields were dominantly from light hadron 
decays, the like-sign subtraction should produce zero pair yield. To 
determine the magnitude of the residual correlation strength after 
like-sign subtraction, the \pp data were fitted with a flat line. This is 
shown as the solid line in Fig.~\ref{fig:punch-throughcorr}a. The fit uncertainty is shown as the 
shaded band around the solid line. The flat fit in \pp had a $\chi^2$/NDF 
of 22.7/24 and gave a value that was nonzero with greater than 1$\sigma$ 
significance. This means there is yield in the final $e$--$\mu$ correlations 
from these punch-through hadrons. The fitted yield was subtracted from the 
final pair yield and its uncertainty was propagated as a systematic 
uncertainty on the final pair yield. For the \dAu case, we fitted the 
residual correlation to a flat line and found reasonable agreement with a 
$\chi^2$/NDF of 30.9/24 or a $p$-value of 14\%.  However, there is a 
possible excess of counts near $\Delta\phi = \pi$, which when included as 
a Gaussian component fixed at $\Delta\phi = \pi$ and the width and yield 
as free parameters, a slightly better $\chi^2$/NDF of 26.3/22 or a 
$p$-value of 26\% was found. If there is any correlated yield beyond a 
pedestal, it would show up in the back-to-back peak. Therefore, we 
subtract the Gaussian fit, shown as the solid line in 
Fig.~\ref{fig:punch-throughcorr} from the final pair yield and propagate 
the uncertainty, shown as the shaded region around the solid line, on the 
fit to the systematic uncertainty in the final pair yield.

Two additional corrections to this data are applied before subtraction 
from the final pair yield. Because the punch-through hadrons are measured in 
the 4th gap, the yields need to be scaled to match the rate of hadrons at 
the last gap. The rate of hadrons at the 5th gap was determined by using 
pion and kaon NLO perturbative QCD spectra~\cite{Vogelsang} and passing them through a 
{\sc geant}-3 model of the PHENIX muon arms. The MuID absorber steel cross 
section was modified until there was agreement between data and the 
simulation for the rate of punch-through hadrons in the 3rd and 4th gap. 
We extrapolated to the 5th gap and find the rate of hadrons is 
2.81$\pm$0.30 times the rate of punch-through hadrons in the 4th 
gap~\cite{Adler:2006yu}. The 3~GeV/$c$ $p_z$ cut removes some fraction of 
the punch-through hadrons. Based on the two-component fit to the $p_z$ 
distribution shown in Fig.~\ref{fig:pz}, the yield is scaled up to account 
for those hadrons rejected by the $p_z$ cut.  In the end, the pair yield 
uncertainty is 2.17$\times$10$^{-9}$ (rad)$^{-1}$ in \pp. In \dAu there is 
a $\Delta\phi$-independent uncertainty on the final pair yield that is 
1.42$\times$10$^{-8}$ (rad)$^{-1}$ and the Gaussian uncertainty that 
ranges from 0 to 6.30$\times$10$^{-8}$ (rad)$^{-1}$.

\subsubsection{Correlations Between Inclusive Electrons and Light-Hadron 
Decay Muons}\label{sec:ppdecaymuon}

One source of real muons is from decays of light hadrons, predominantly 
charged pions and kaons, before the absorber material. The observed rate 
of muons into the North arm is higher, when the collision vertex is farther 
from the spectrometer arm. Because heavy flavor decays (including Drell-Yan, 
heavy quarkonia, etc.)~have a much shorter $c\tau$ than light flavor 
decays, heavy flavor decay muons have a much weaker vertex dependence. 
Therefore, we assume there are two components to the muon rate: a 
component that follows the primary vertex distribution, attributable to 
heavy flavor decays, and a component that folds the linear component due 
to light hadron decays with the primary vertex distribution.

\begin{figure}[thb]
\includegraphics[width=1.0\linewidth]{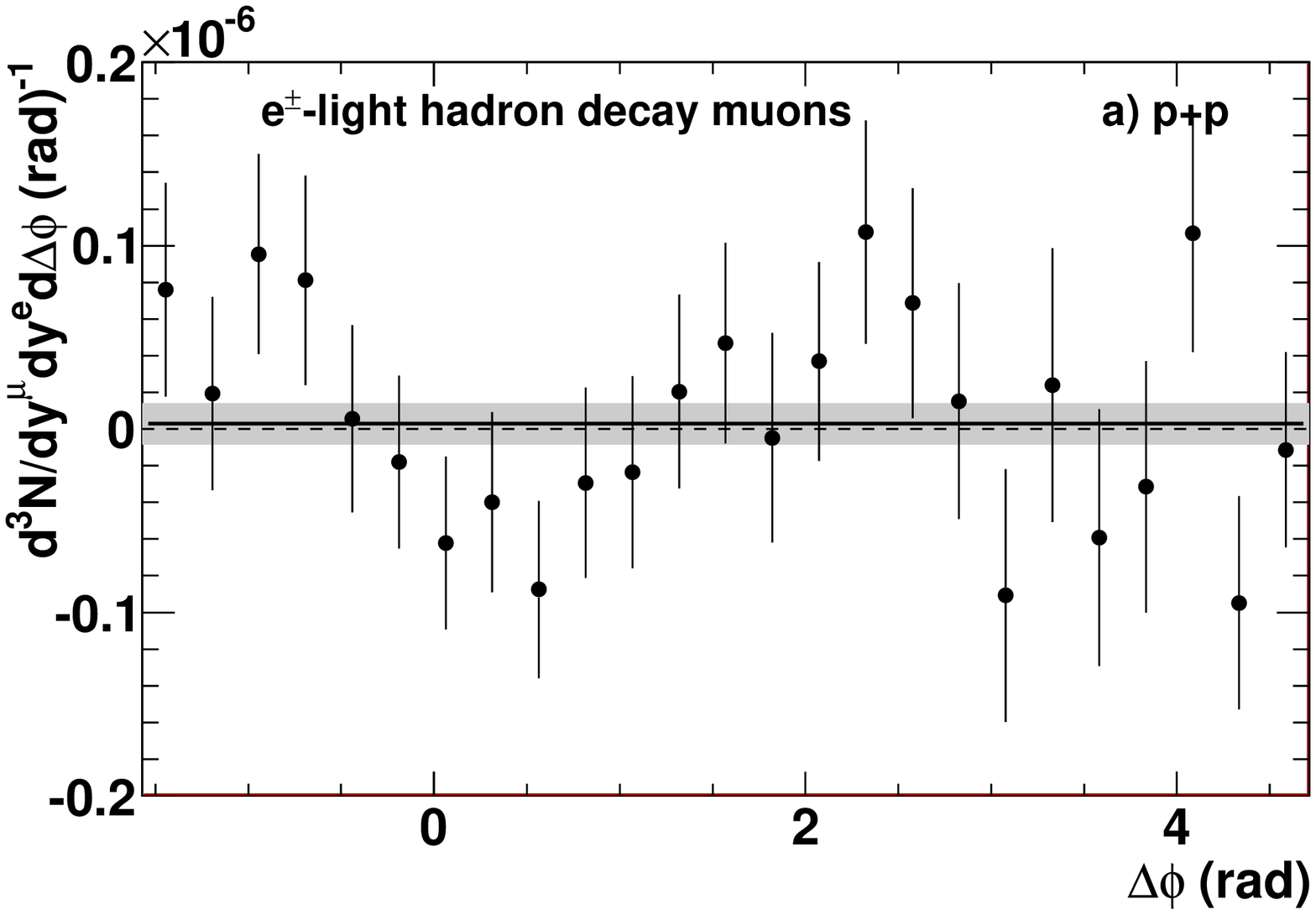}
\includegraphics[width=1.0\linewidth]{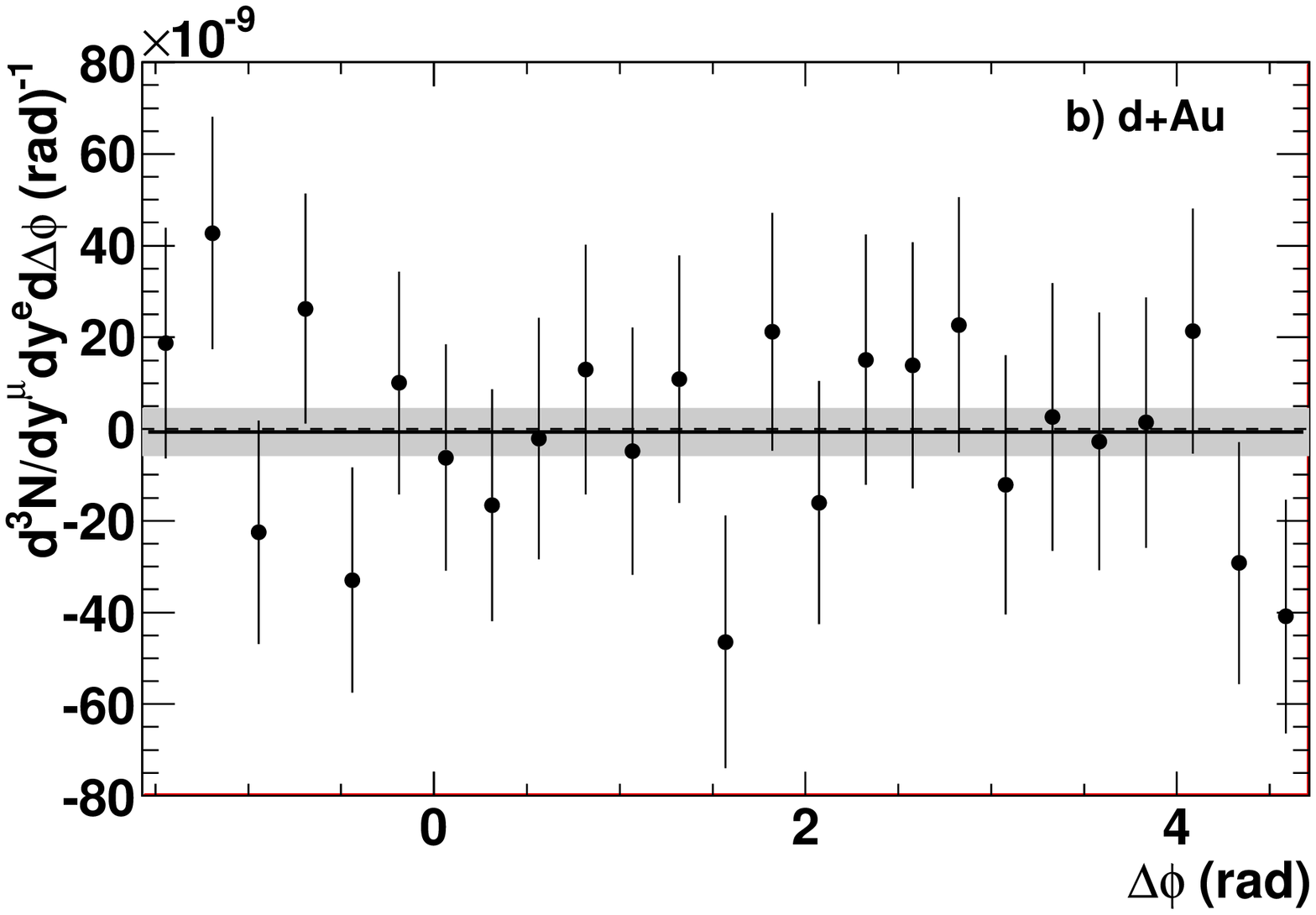}
\caption{
The fully-corrected like-sign-subtracted and near-far vertex-subtracted 
(see text) muon-decay $\Delta\phi$ pair yield in (a) \pp and (b) \dAu 
collisions. Both are consistent with no residual correlation after 
like-sign subtraction. The solid lines and shaded bands indicate the flat 
line fits and their uncertainty with $\chi^2$/NDF of 27.1/24 and 18.0/24 
in \pp and~\dAu, respectively.
}
\label{fig:decaycorr}
\end{figure}

Muons that are near the detector ($0<z_{\rm vtx}<30$~cm) and far from 
the 
detector ($-30<z_{\rm vtx}<0$~cm), where $z_{\rm vtx}$ is the measured 
collisions 
vertex, are separately correlated with central arm electrons. Because the 
signal heavy flavor muons follow the primary collision vertex 
distribution, subtracting the near-vertex pair yield from the far-vertex 
pair yield, should remove these and only residual correlations from decay 
muons should be present. The pair yields in \pp and \dAu after subtracting 
near- and far-vertex muons and after like-sign subtraction are shown in 
Fig.~\ref{fig:decaycorr}. The \dAu correlations are consistent with a flat 
line with zero yield with a $\chi^2$/NDF of 18.0/24. The \pp data seems to 
have a residual shape. However, this shape is asymmetric about 
$\Delta\phi$ of zero and is not physical. Therefore, we fit with a flat 
line that results in zero correlation yield and a $\chi^2$/NDF of 27.1/24. 
The fits are shown in Fig.~\ref{fig:decaycorr} as solid lines and shaded 
bands indicating the statistical uncertainties. These uncertainties were 
propagated into the systematic uncertainties of the final pair yields.

To propagate the uncertainties, additional corrections are 
needed. First, in the far-near subtraction, some fraction of the decay 
muons are removed. Second, light hadron decays outside the $\pm$30~cm 
vertex cut are not counted in the subtraction. To account for both 
effects, a fit to the vertex dependence of the muon yield is extrapolated 
to a point one interaction length inside the absorber, a distance of about 
56~cm from the nominal $z$ vertex and about 16~cm into the absorber. It is 
assumed that the decay contribution to the muons is negligible at that 
point, which fixes the fraction of muons that are from light decays within 
the measured vertex window of the analysis. Under this assumption, only 
22\% of the decay muons are measured within the vertex window after the 
like-sign subtraction. The fit uncertainties are increased to account for 
those muons not measured. The final systematic uncertainties on the final 
pair yield are 1.13$\times$10$^{-8}$ (rad)$^{-1}$ and 
5.05$\times$10$^{-8}$, independent of $\Delta\phi$ for \pp and \dAu, 
respectively.

\subsubsection{Correlations Between Photonic Electrons and Inclusive 
Muons}\label{sec:photonic}

Electrons can result from light hadron decays through internal and 
external photon conversions. The dominant photonic source of electrons are 
from $\pi^0$ decays. We assume that, if we measure the $\pi^0$-decay 
electrons correlations with muons, this will represent the other photonic 
sources (such as $\eta$ and $\omega$ decay) in shape and yield. To tag 
decay or converted electrons, we construct the invariant mass distribution 
of all pairs of electrons and photons in an event. Electrons paired with 
photons within the $\pi^0$ mass peak are then correlated with muon 
candidates. The signal-to-background of pairs in the $\pi^0$ mass range is 
about one. To remove correlations from combinatorial electron-photon pairs 
that fall within the $\pi^0$ mass window, muon candidates were also 
correlated with the $e$--$\gamma$ pairs in a ``sideband'' $\pi^0$ mass 
region from 0.2--0.4~GeV/$c^2$.  After scaling by the appropriate 
signal-to-background under the $\pi^0$ mass region, the ``sideband'' 
correlations were subtracted from the in-mass electron-muon correlations 
for each of the $e$--$\mu$ charge types.

\begin{figure}[thb]
\includegraphics[width=1.0\linewidth]{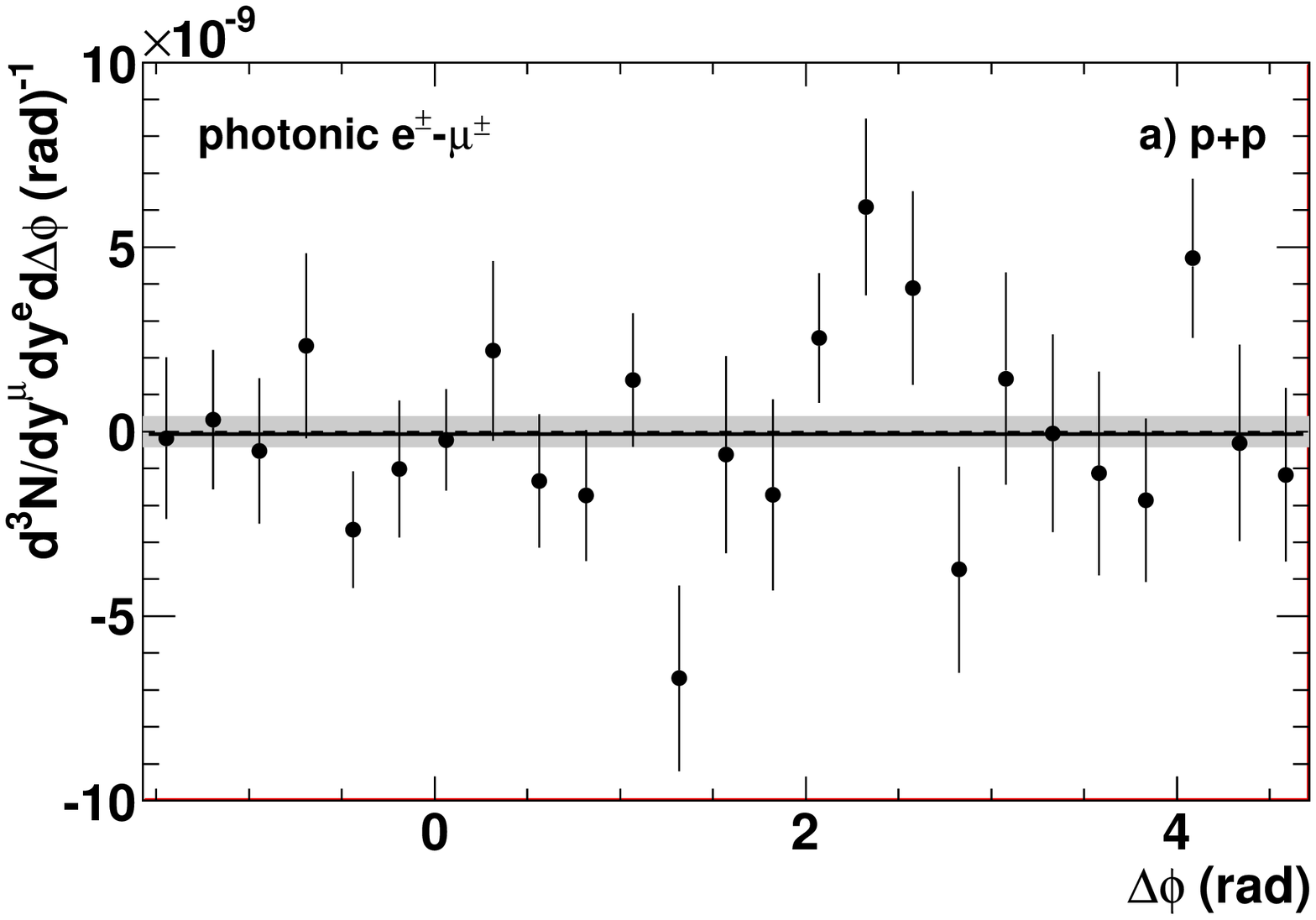}
\includegraphics[width=1.0\linewidth]{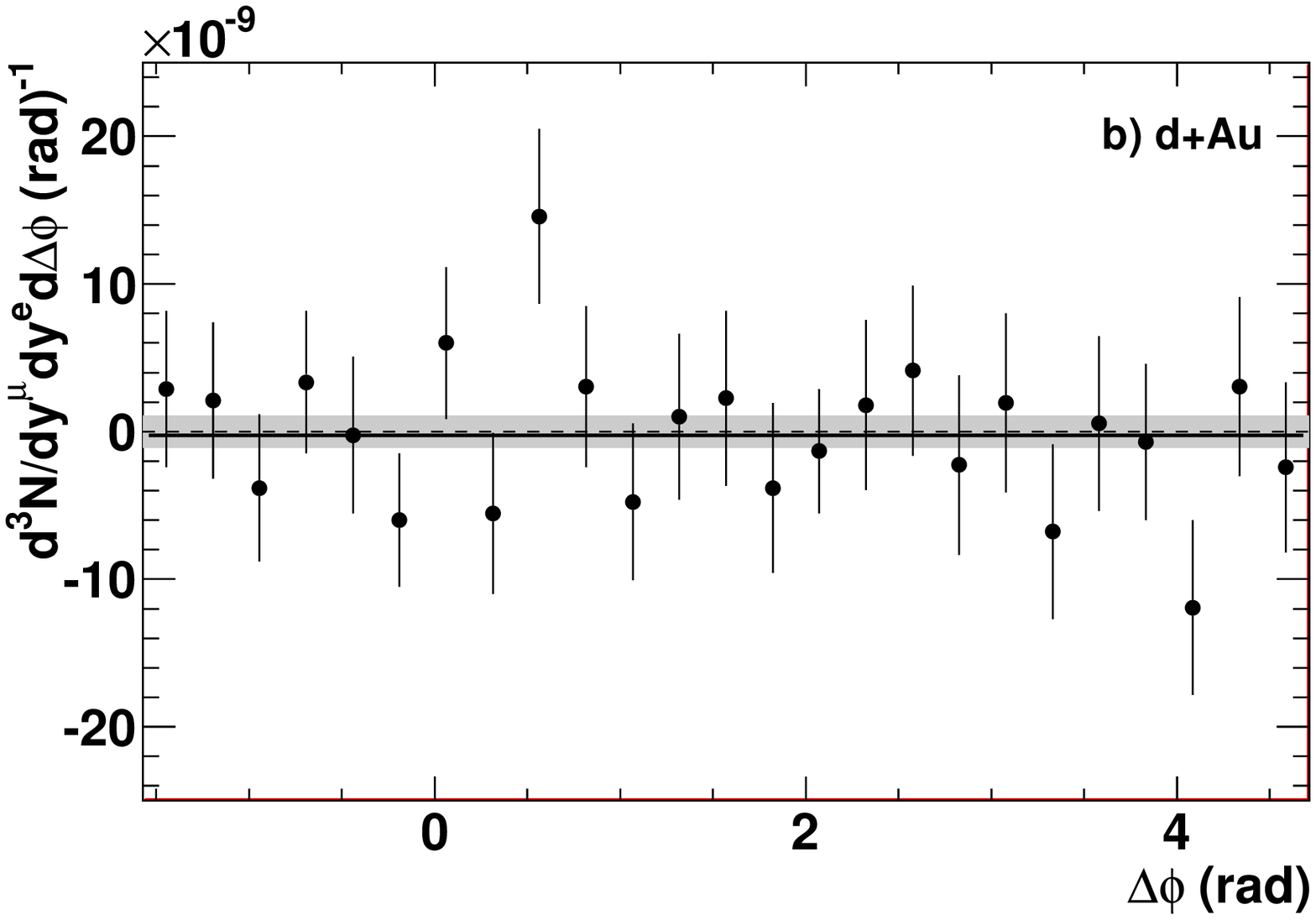}
\caption{
The fully-corrected like-sign-subtracted photonic electron-muon 
$\Delta\phi$ pair yield in (a) \pp and (b) \dAu collisions. Both are 
consistent with no residual correlation after like-sign subtraction. The 
solid lines and shaded bands indicate the flat line fits and their 
uncertainties with $\chi^2$/NDF of 33.2/24 and 20.2/24 in \pp and~\dAu, 
respectively.
}
\label{fig:photoniccorr}
\end{figure}

Fig.~\ref{fig:photoniccorr} shows the ``sideband''-subtracted and 
like-sign subtracted correlation between electrons tagged in the $\pi^0$ 
mass region with muons from \pp and \dAu data. Flat fits to these 
correlations produced a yield consistent with zero with $\chi^2$/NDF of 
33.2/24 and 20.2/24 in \pp and \dAu data, respectively. The statistical 
uncertainty from the fitted yield to the $\pi^0$-tagged correlations is a 
factor of 10 smaller than the other background correlations after 
accounting for reconstruction efficiency and additional sources of 
photonic electrons. This uncertainty is negligible compared to those from 
the muon backgrounds.

\subsection{Systematic Uncertainties}\label{sec:syserr}

In this analysis there are three general types of uncertainty that we 
identify as type A, point-to-point uncorrelated, type B, point-to-point 
but correlated, and type C, total normalization uncertainty. Except for 
statistical uncertainties there are no type A uncertainties in this 
analysis.

The type B uncertainties are from the subtraction of known backgrounds 
discussed in Section~\ref{sec:subtr}. The fully corrected pair yield 
uncertainties in \pp are $2.17\times10^{-9}$ (rad)$^{-1}$ and 
$1.13\times10^{-8}$ (rad)$^{-1}$ from punch-through hadron and decay 
hadron subtraction uncertainties, respectively. These values are 
independent of $\Delta\phi$. In \dAu the flat-line fit contributions to 
the systematic uncertainty are $1.42\times10^{-8}$ (rad)$^{-1}$ and 
$5.05\times10^{-8}$ (rad)$^{-1}$ from punch-through hadron and decay 
hadron subtraction uncertainties, respectively. The additional uncertainty 
from the Gaussian fit to the punch-through hadron correlations resulted in 
a $\Delta\phi$-dependent uncertainty ranging in absolute value of 0 at 
$\Delta\phi\sim$ 2 rad to $6.30\times10^{-8}$ (rad)$^{-1}$ at 
$\Delta\phi\sim\pi$. The type-B systematics are summarized in 
Table~\ref{tab:sys}.

\begingroup
\squeezetable
\begin{table}
\caption{
Table of type B and type C systematic uncertainties for \pp and \dAu 
collision data. The uncertainties on the muon and electron cuts are highly 
correlated between \pp and \dAu.
}
\label{tab:sys}
\begin{ruledtabular} \begin{tabular}{cccc}
Type & Description & \pp & \dAu \\
\hline
B & $\Delta\phi$ dependent & -- & 0\%---6.30$\times$10$^{-8}$(rad)$^{-1}$ \\
B & punch-through & $2.17\times10^{-9}$ (rad)$^{-1}$ & $1.42\times10^{-8}$(rad)$^{-1}$ \\
B & decay muons & $1.13\times10^{-8}$ (rad)$^{-1}$ & $5.05\times10^{-8}$(rad)$^{-1}$ \\
\\
C & muon cuts & 7.8\% & 8.3\% \\
C & electron cuts & 8.3\% & 9.3\% \\
C & muon efficiency & 2.2\% & 2.2\% \\
C & electron efficiency & 1.0\% & 1.0\% \\
C & trigger efficiency & 11.1\% & 4.2\% \\
\\
C & total & 16.1\% & 13.4\% \\
\end{tabular} \end{ruledtabular}
\end{table}
\endgroup

The type C uncertainties are attributable to several sources and are given 
in Table~\ref{tab:sys}. One source of systematic uncertainty is evaluated 
by tightening the single particle cuts for this analysis. Each single 
particle cut was tightened independently and the analysis, including 
reevaluation of the single particle efficiency, was performed. The 
uncertainty from each of the individual single particle cuts was combined 
using the correlation amongst the cuts. The values of these are different 
in \pp and \dAu data, because of the higher backgrounds in \dAu 
collisions. However, these uncertainties are highly correlated between \pp 
and \dAu, because the same cuts are applied to both data sets. Another 
source of uncertainty is in the evaluation of the single particle 
efficiencies. The single particles were generated flat in $p_T$ and then 
weighted to match the measured PHENIX heavy flavor lepton 
spectra~\cite{Adare:2006hc}. For the uncertainty determination, the single 
particle efficiency was re-evaluated without the weighting applied. This 
was estimated to be 1.0\% for the electrons and 0.8\% for the muons. For 
muons there is an additional 2.0\% uncertainty due to the run-by-run 
variation in muon acceptance. The final portion of the type C systematic 
uncertainty is due to the trigger efficiency. To evaluate this 
uncertainty, the data were analyzed for several data-taking periods 
defined by the muon trigger performance. The difference in fully-corrected 
yields between data sets was taken to be the uncertainty in the muon 
trigger efficiency. This is combined with the uncertainties in the bias 
factor $c$ in Eq.~\ref{eq:pairyield}.  The total uncertainty for the 
trigger is 11.1\% for \pp and 4.2\% for \dAu.  As indicated in 
Table~\ref{tab:sys}, combining all Type C uncertainties gives 16.1\% for 
\pp data and 13.4\% for \dAu data.

\section{Results}\label{sec:results}

\subsection{Pair Yields for \pp data and Comparison with Monte Carlo 
Generators}\label{sec:yields}

\begin{figure}[thb]
\includegraphics[width=1.0\linewidth]{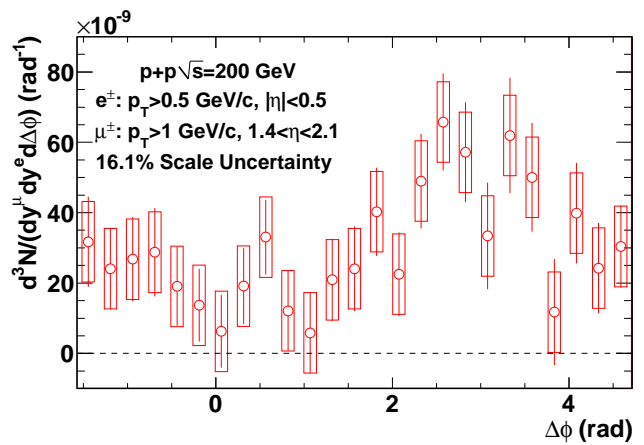}
\caption{(Color Online) 
The fully-corrected like-sign-subtracted heavy flavor $e$--$\mu$ pair 
yield in \pp. The error bars are statistical only. The boxes show the type B 
systematic uncertainty from the punch-through hadron and light hadron 
decay muon background subtraction. The 16.1\% type C systematic 
uncertainty is not shown.
}
\label{fig:ppcorr}
\end{figure}

The fully-corrected like-sign subtracted $e$--$\mu$ pair yield as a function 
of $\Delta\phi$ for electrons with $p_T>0.5$~GeV/$c$ and $|\eta|<0.5$, 
with opposite-signed forward muons with $p_T>1.0$~GeV/$c$ and 
$1.4<\eta<2.1$, in \pp is shown in Fig.~\ref{fig:ppcorr}. The average muon 
$\eta$ in these correlations is 1.75. The error bars are statistical 
uncertainties only, while the boxes are the type B systematic 
uncertainties. We note that the distribution has two components: a nonzero 
continuum as well as a back-to-back peak near $\Delta\phi=\pi$.

\begin{figure}[thb]
\includegraphics[width=1.0\linewidth]{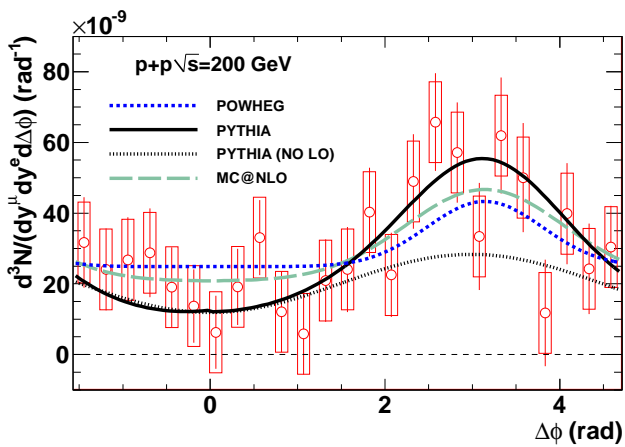}
\caption{(Color online) 
Comparison of the measured \pp pair yield ([red] points) with heavy flavor 
production in {\sc powheg} ([blue] dashed line), {\sc pythia} ([black] 
solid line) and MC@NLO ([green] long dashed line). The $e$--$\mu$ pair 
yield from the subset of {\sc pythia} events, when the $c\bar{c}$ is not 
produced at the event vertex is plotted as the dotted [black] line. Each 
Monte Carlo curve was scaled by a single parameter to match the observed 
yield. The resulting cross sections are consistent with the previously 
measured PHENIX results (see Table~\ref{tab:xsect}).
} 
\label{fig:compare}

\end{figure}

To interpret these data, we compare the \pp results to several different 
Monte Carlo generators, {\sc pythia}, {\sc powheg}\cite{Frixione:2007nw}, 
and MC@NLO\cite{Frixione:2002ik}.

The {\sc pythia} MB QCD events were generated to model the LO 
gluon fusion process and also model next-to-leading order processes, like 
flavor excitation and gluon splitting. Events with a $c\bar{c}$ pair and 
an electron and a muon in the measured kinematic range as the corrected 
data ($p_T^e>0.5$~GeV/$c$ and $|\eta^e|<0.5$, $p_T^\mu>1$~GeV/$c$ and 
$1.4<\eta^\mu<2.1$) were correlated and a like-sign subtraction was 
performed. An overall scale factor was used to fit the {\sc pythia} curve 
to the \pp data. In the fit, the $\chi^2$ was calculated for different 
scale parameters using the statistical error on the \pp data. We report 
the cross section for the scale factor that minimizes that $\chi^2$ and 
report a statistical error on the cross section as the value that changes 
the $\chi^2$ by one unit. To evaluate the systematic uncertainty on the 
cross section, the \pp data were increased and decreased by their combined 
type B and type C systematic uncertainty and the process to determine the 
scale factor by finding a minimum $\chi^2$ using the statistical 
uncertainty in the data was repeated. We find the {\sc pythia} correlation 
is consistent with the \pp data with a $c\bar{c}$ cross section of 
$\sigma_{c\bar{c}}$ = 340$\pm$29(stat)$\pm$116(syst)~$\mu$b with a 
$\chi^2$/NDF of 20.5/24. This is shown as the solid curve in 
Fig.~\ref{fig:compare}.

The other model comparisons are from NLO generators, {\sc powheg} and 
MC@NLO. Events were generated to produce the hard scattering heavy flavor 
event vertex and then interfaced to {\sc pythia}, which performed the 
fragmentation and underlying event generation. The qualitative features of 
the data are present in these correlations: the continuum and the 
back-to-back peak. As described for the {\sc pythia} fit, a single scale 
parameter was used to calculate a $\chi^2$ between the generated 
$e$--$\mu$ correlations and the data using the data's statistical 
uncertainty. The resulting best fits for {\sc powheg} and MC@NLO are shown 
in Fig.~\ref{fig:compare} as the short dashed and the long dashed lines, 
respectively. The extracted cross sections are $\sigma_{c\bar{c}}$ = 511 
$\pm$ 44 (stat)~$\pm$ 198 (syst)~$\mu$b with $\chi^2$/NDF of 23.5/24 for 
{\sc powheg} and $\sigma_{c\bar{c}}$ = 764 $\pm$ 64 (stat)~$\pm$ 284 
(syst)~$\mu$b with $\chi^2$/NDF of 19.2/24 for MC@NLO.

We combine the cross sections from the three models and report a measured 
cross section of $\sigma_{c\bar{c}} =$ 538 $\pm$ 46 (stat) $\pm$ 197 
(data~syst) $\pm$ 174 (model~syst). The central value of the cross section 
is the average of the three model cross sections, while the model 
systematic uncertainty is the standard deviation of the three model cross 
sections. This value can be compared with previous PHENIX measurements. 
From the heavy flavor electron spectra at midrapidity, PHENIX found 
$\sigma_{c\bar{c}} =$ 567 $\pm$ 57 (stat) $\pm$ 224 
(syst)~\cite{Adare:2006hc} and from the dielectron mass spectrum at 
midrapidity, PHENIX extracted $\sigma_{c\bar{c}} =$ 554 $\pm$ 39(stat) 
$\pm$ 142~(data~syst)~$\pm$ 200~(model~syst)~\cite{Adare:2008qb}. Within 
the data systematics the value extracted here is consistent with 
previously published PHENIX results.

\begin{table*}
\caption{Table of measured $c\bar{c}$ cross sections from previous PHENIX 
analysis and from Monte Carlo generators compared to the $e$--$\mu$
correlations in this analysis.
}
\begin{ruledtabular} \begin{tabular}{cccc}
& description & $\sigma_{c\bar{c}}$ ($\mu$b) & \\
\hline
& {\sc pythia} $e$--$\mu$ & 340$\pm$29(stat)$\pm$116(syst) & \\
& {\sc powheg} $e$--$\mu$ & 511$\pm$44(stat)$\pm$198(syst) & \\
& MC@NLO $e$--$\mu$ & 764$\pm$64(stat)$\pm$284(syst) & \\
& Combined $e$--$\mu$ & 538$\pm$46(stat)$\pm$197(data syst)$\pm$174(model syst)  & \\
\\
& PHENIX single $e^\pm$~\protect\cite{Adare:2006hc} 
	& 567$\pm$57(stat)$\pm$224(syst) & \\
\\
& PHENIX dilepton ($e^+e^-$)~\protect\cite{Adare:2008qb} 
   & 554$\pm$39(stat)$\pm$142(data syst)$\pm$200(model syst) & \\
\end{tabular} \end{ruledtabular}
\label{tab:xsect}
\end{table*}

Using the {\sc pythia} event record, it is possible to separate the 
$c\bar{c}$ production into an LO component, where the 
$gg(q\bar{q})\rightarrow c\bar{c}$ and a component from the {\sc pythia} 
model of NLO mechanisms of flavor excitation and gluon splitting, where 
the $c\bar{c}$ pair is produced in the initial or final-state shower. The 
``{\sc pythia} (NO LO)'' dashed line in Fig.~\ref{fig:compare} shows the 
correlations from the sample of produced {\sc pythia} events, where the 
$c\bar{c}$ were not generated in the primary event vertex of {\sc pythia}. 
The back-to-back peak at $\Delta\phi = \pi$ is dominated by the LO gluon 
fusion process while the continuum is due to the correlations from the 
higher order processes. From an accounting from {\sc pythia}, we find that 
32\% of the $e$--$\mu$ pair yield results from gluon fusion, consistent 
with the expectations from charm production~\cite{Brambilla:2010cs}.

Throughout the analysis it has been assumed that semileptonic $c\bar{c}$ 
decay is the dominant contribution to the correlations. However, 
$b\bar{b}$ semileptonic decays would produce a signal in both the like- 
and the unlike-sign pair distributions. Up to four semileptonic decays can 
occur where $b$-quarks semileptonically decay to $c$-quarks, which 
subsequently semileptonically decay. We have used {\sc pythia} and {\sc 
powheg} to check these contribution from bottom. In both cases, for 
electrons and muons in the kinematic region that we measure, the bottom 
contribution is about a factor of 100 below the charm yield. This is 
further corroborated by the PHENIX heavy flavor electron measurements that 
show that bottom becomes significant only at $p_T$ above 3 
GeV/$c$~\cite{Adare:2009ic}. In this analysis only 3\% of the sampled 
electrons have a $p_T$ above 3~GeV/$c$, so we expect that the bottom 
contribution is negligible in this measurement especially compared to the 
background subtraction systematic uncertainties.

\subsection{Yields in $d$+${\rm Au}$ and Comparison to \pp}
\label{sec:dAuresults}

\begin{figure}
\includegraphics[width=1.0\linewidth]{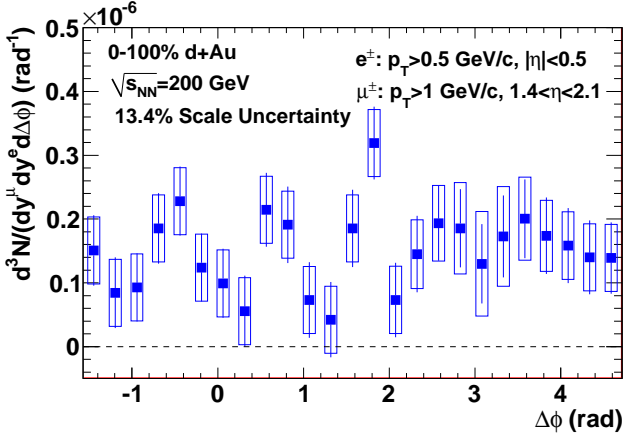}
\caption{(Color Online) 
The fully corrected like-sign-subtracted heavy flavor $e$--$\mu$ pair yield 
in \dAu. The error bars are statistical only. The boxes show the type B 
systematic uncertainty from the punch-through hadron and light hadron 
decay muon background subtraction. The 13.4\% type C systematic 
uncertainty is not shown.
}
\label{fig:dAucorr}
\end{figure}

\begin{figure}[thb]
\includegraphics[width=1.0\linewidth]{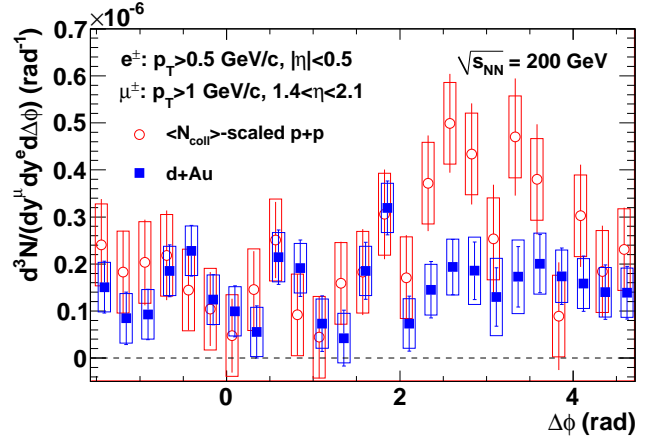}
\caption{(Color online) 
The fully-corrected like-sign-subtracted heavy flavor $e$--$\mu$ pair yield 
in ([red] circles) $\langle N_{\rm coll}\rangle$-scaled \pp ([blue] boxes) 
\dAu, shifted in $\Delta\phi$ for clarity. The bars are statistical 
uncertainty. The boxes are the type B systematic uncertainty from the 
decay and punch-through background subtraction. The overall normalization 
uncertainties of 16.1\% and 13.4\% in \pp and \dAu, respectively, and 
5.7\% uncertainty from $N_{\rm coll}$ are not included.
}
\label{fig:paircorr}
\end{figure}

The fully-corrected like-sign subtracted pair yield as a function of 
$\Delta\phi$ for electrons with $p_T^e>0.5$~GeV/$c$ and $|\eta^e|<0.5$ 
with forward muons with $p_T^\mu>1.0$~GeV/$c$ and $1.4<\eta^\mu<2.1$ in 
0\%--100\% \dAu, corresponding to the total inelastic cross section, is 
shown in Fig.~\ref{fig:dAucorr}.  A nonzero correlations strength 
is observed. 
However, unlike the \pp data, there is a much less distinct back-to-back 
peak near $\Delta\phi$ of $\pi$. Fig.~\ref{fig:paircorr} shows the overlay 
of the \pp and \dAu pair correlations. The \pp pair correlations are  
scaled by the \dAu 
$\langle N_{\rm coll}\rangle = 7.59\pm0.43$~\cite{Adare:2013nff}. The peak 
in \dAu is suppressed compared to \pp, indicating a medium modification to 
the yield per collision in \dAu.

\begin{figure}
\includegraphics[width=1.0\linewidth]{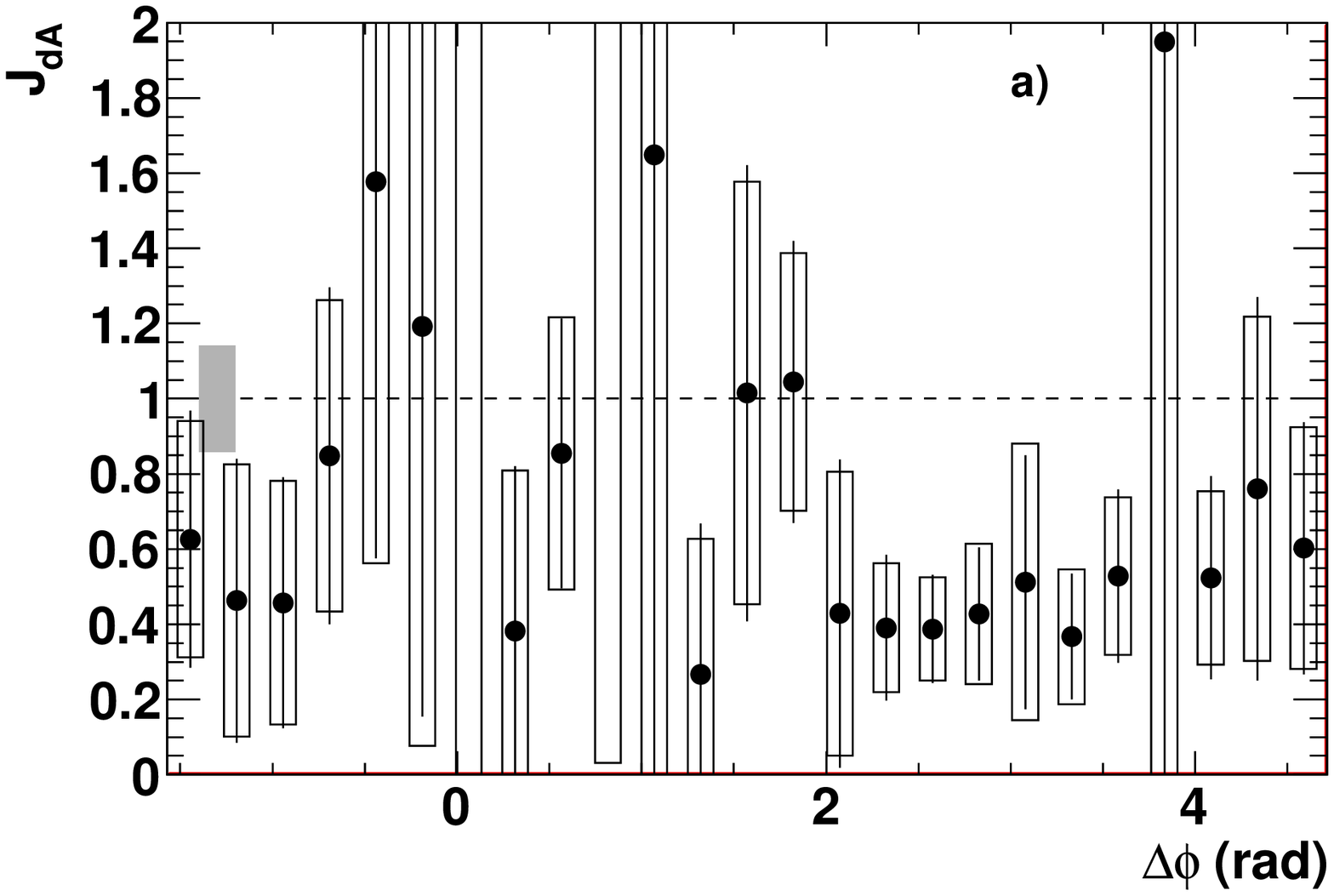}
\includegraphics[width=1.0\linewidth]{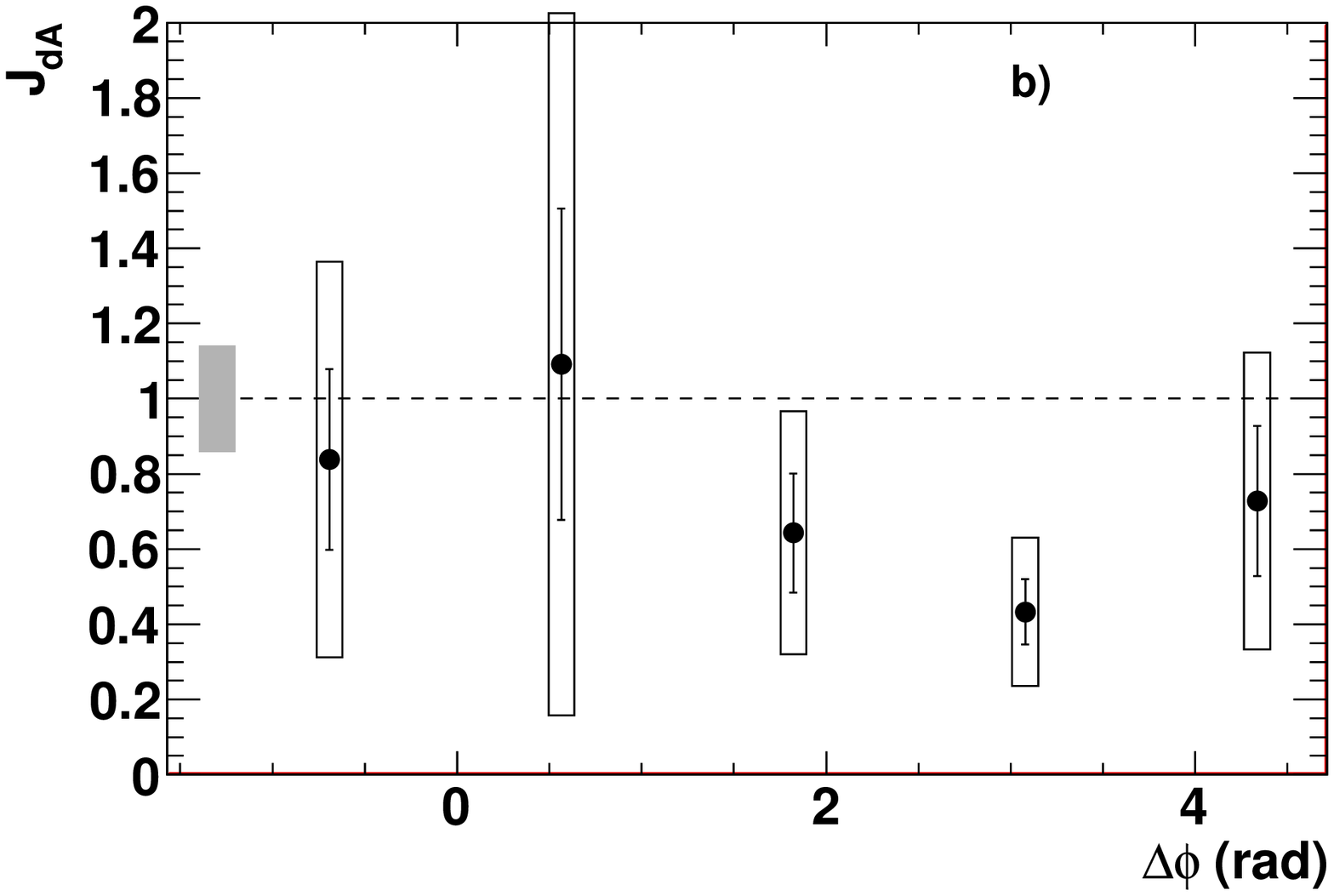}
\caption{
$J_{dA}$ is plotted as a function of $\Delta\phi$.  The vertical bars are 
statistical uncertainties, the black boxes are the type B systematic 
uncertainties, and the gray band around 1.0 on the left is the type C 
systematic uncertainty. The type B systematics are symmetric around the 
central value but in some cases are outside the range of the plot. 
(b) alternative rebinning of the data in (a).
}
\label{fig:jdadphi}
\end{figure}

To quantify the difference between \pp and \dAu yields, we calculate the 
ratio $J_{dA}$ defined as the ratio of a pair yield in \dAu to the 
$N_{\rm coll}$-scaled pair yield in \pp.
\begin{equation}\label{eq:jda}
J_{dA} = \frac{d+{\rm Au~pair~yield}}{\langle N_{\rm coll}\rangle~p+p~{\rm pair~yield}}.
\end{equation}
Any deviation from unity of this ratio would indicate modification to the 
yield. When taking this ratio several systematic uncertainties common to 
\pp and \dAu cancel. These are dominantly from identical cuts used in the 
analyses with the same systematic uncertainties. The noncanceling type C 
systematic uncertainties in the \pp and \dAu yields are 7.7\% and 8.9\%, 
respectively.

Fig.~\ref{fig:jdadphi}a shows a plot of $J_{dA}$ as a function of 
$\Delta\phi$ for all bins in $\Delta\phi$. The bars are statistical 
uncertainties and the type B systematic uncertainties are plotted as 
boxes. The noncanceling type-C uncertainty is 14.1\% and is indicated by 
the shaded box around one on the left. While the points near $\Delta\phi = 
0$ are consistent with unity with large error bars, the points near 
$\Delta\phi = \pi$, where $J_{dA}$ is about 0.4. 
For clarity, Fig.~\ref{fig:jdadphi}b shows a rebinning of 
Fig.~\ref{fig:jdadphi}a.  We find
\begin{eqnarray}
J_{dA}(2.7<\Delta\phi<3.5{\rm ~rad}) & = & 0.433 \pm 0.087 {\rm (stat)} \nonumber \\
& & \pm 0.135 {\rm (syst})
\end{eqnarray}
for the bin near $\Delta\phi = \pi$. This value is 3.5$\sigma$ different 
from unity after combining the statistical and systematic uncertainties.

These results show that, in the measured kinematics, charm pairs are 
modified in the cold nuclear medium. These results are in a different 
kinematic region than either the single electrons, which are enhanced at 
midrapidity~\cite{Adare:2012qb}, or the single muons, which are suppressed 
at forward rapidity~\cite{Adare:2013lkk}. From the {\sc pythia} 
simulation, the $e$--$\mu$ correlations arise from partons in the gold 
nucleus with $x\approx10^{-2}$ at $Q^{2}\approx10~{\rm GeV}^{2}$, on the 
edge of the shadowing region. As discussed in Section~\ref{sec:yields}, 
the back-to-back peak is dominated by leading order gluon fusion, while 
the continuum is dominated by other processes like flavor excitation and 
gluon splitting. The observed back-to-back peak and pedestal in \pp and 
\dAu should help lead to an understanding of the mechanism or mechanisms 
responsible for the modification. For example, the back-to-back peak is 
dominated by low-$x$ gluons participating in the hard scattering, whereas 
the continuum has a larger contribution of quarks participating in the 
hard scattering. Quarks are probably less shadowed than gluons at the $x$ 
and $Q^2$ where this analysis is measured. It is possible that there are 
kinematic differences between the final state charm quarks in the peak and 
the continuum. These differences could affect the amount of final state 
energy loss and multiple scattering that modify the measured pair yields. 
It may be possible to combine these results with other cold nuclear matter 
charm measurements to disentangle the effects of shadowing, saturation, 
and energy loss.

\section{Summary and conclusions}\label{sec:end}

We presented PHENIX results for heavy flavor production of 
azimuthally-correlated unlike sign $e$--$\mu$ pairs in \pp and 
\dAu collisions at $\sqrt{s_{NN}}$ of 200~GeV. The \pp yield shows a 
nonzero continuum as well as a back-to-back peak structure centered at 
$\Delta\phi=\pi$. When compared with several models, we find the charm 
cross section $\sigma_{c\bar{c}} =$ 538 $\pm$ 46 (stat) $\pm$ 197 
(data~syst) $\pm$ 174 (model~syst) $\mu$b. This is also consistent with 
previously measured $c\bar{c}$ cross sections at this center of mass 
energy. In \dAu collisions a yield reduction in the back-to-back peak is 
observed, where we measure $J_{dA}(2.7<\Delta\phi<3.5{\rm ~rad})$ = 
0.433 $\pm$ 0.087 (stat) $\pm$ 0.135 (syst). This indicates that the 
nuclear medium modifies the $c\bar{c}$ correlations. Such a suppression 
could arise due to nuclear PDF shadowing, saturation of the gluon 
wavefunction in the Au nucleus, or initial/final state energy loss and 
multiple scattering.


\section*{ACKNOWLEDGMENTS}


We thank the staff of the Collider-Accelerator and Physics
Departments at Brookhaven National Laboratory and the staff of
the other PHENIX participating institutions for their vital
contributions.  We acknowledge support from the 
Office of Nuclear Physics in the
Office of Science of the Department of Energy,
the National Science Foundation, 
a sponsored research grant from Renaissance Technologies LLC, 
Abilene Christian University Research Council, 
Research Foundation of SUNY, and
Dean of the College of Arts and Sciences, Vanderbilt University 
(U.S.A),
Ministry of Education, Culture, Sports, Science, and Technology
and the Japan Society for the Promotion of Science (Japan),
Conselho Nacional de Desenvolvimento Cient\'{\i}fico e
Tecnol{\'o}gico and Funda\c c{\~a}o de Amparo {\`a} Pesquisa do
Estado de S{\~a}o Paulo (Brazil),
Natural Science Foundation of China (P.~R.~China),
Ministry of Education, Youth and Sports (Czech Republic),
Centre National de la Recherche Scientifique, Commissariat
{\`a} l'{\'E}nergie Atomique, and Institut National de Physique
Nucl{\'e}aire et de Physique des Particules (France),
Bundesministerium f\"ur Bildung und Forschung, Deutscher
Akademischer Austausch Dienst, and Alexander von Humboldt Stiftung (Germany),
Hungarian National Science Fund, OTKA (Hungary), 
Department of Atomic Energy and Department of Science and Technology (India),
Israel Science Foundation (Israel), 
National Research Foundation and WCU program of the 
Ministry Education Science and Technology (Korea),
Physics Department, Lahore University of Management Sciences (Pakistan),
Ministry of Education and Science, Russian Academy of Sciences,
Federal Agency of Atomic Energy (Russia),
VR and Wallenberg Foundation (Sweden), 
the U.S. Civilian Research and Development Foundation for the
Independent States of the Former Soviet Union, 
the US-Hungarian Fulbright Foundation for Educational Exchange,
and the US-Israel Binational Science Foundation.



\end{document}